\documentclass[aps,twocolumn,floatfix]{revtex4}
\usepackage{graphicx} 
\usepackage{amsmath}
\usepackage{color}
\newcommand{\bea}{\begin{eqnarray}}
\newcommand{\eea}{\end{eqnarray}}
\newcommand{\bse}{\begin{subequations}}
\newcommand{\ese}{\end{subequations}}

\begin{document}

\title{Thermodynamics of the nonrelativistic free-electron Fermi gas in one, two, and three dimensions from the degenerate to the nondegenerate temperature regime}

\author{David C. Johnston}
\affiliation{Ames Laboratory and Department of Physics and Astronomy, Iowa State University, Ames, Iowa 50011, USA}

\date{\today}

\begin{abstract}

The thermodynamic properties of a nonrelativistic free-electron Fermi gas is of fundamental interest in condensed matter physics.  Properties previously studied in three-dimensions (3D) in the low- and high-temperature limits include the internal energy, heat capacity, zero-field magnetic spin susceptibility, and pressure.   Here we report solutions for the temperature dependence  spanning these two temperature regimes of the chemical potential, internal energy, magnetic susceptibility, and the heat capacity at constant volume in 1D, 2D, and 3D\@.  Also calculated are the pressure, enthalpy, heat capacity at constant pressure, isothermal compressibility, and thermal expansion coefficient versus temperature in 2D and 3D\@.   Of primary interest here are the detailed dimension-dependent crossovers of these properties between the degenerate and nondegenerate temperature regime, which are graphically illustrated for each of the above properties.

\end{abstract}

\maketitle

\section{Introduction}

A free-electron Fermi gas as defined here is a collection of a large number $N$ of nonrelativistic noninteracting electrons with spin $S = 1/2$ in zero external potential.  Thus the interaction of the electrons with a crystal lattice of positive ions that would give rise to energy gaps in the band structure and possibly superconductivity is ignored.  The electron gas can be in one dimension (1D), 2D, or 3D\@.  The low- and high-temperature~$T$ limits of the thermal and magnetic properties of a 3D free-electron Fermi gas are well known~\cite{Sommerfeld1928, Ashcroft1976, Kittel1980, Ibach2002, Kittel2005, Hook2010, Simon2013, Grosso2014}.  Other aspects and consequences of the Fermi statistics have also been considered~\cite{McKelvey1964, Kapron1970, Bludman1977, Arnaud1999, Balian1999, Mulin2003, Silbar2004, Durand2004, Nattermann2005, Muller2015, Cetina1977, Sevilla2016, Li1998, Johnston2020}.  In the limit of high $T$ the heat capacity at constant volume $C_{\rm V}$ and the pressure $p$ are the same as for a monatomic  ideal gas and the magnetic spin susceptibility $\chi$ is the same as for $N$ isolated electrons following the Curie law $C/T$, where $C$ is the Curie constant for $S=1/2$.  At low $T$ in the degenerate regime, $C_{\rm V}$ is proportional to $T$, and $\chi$ and $p$ saturate to constant values as $T \to 0$~K\@.  Here expressions for the chemical potential $\mu(T)$ are discussed~\cite{Cowan2019} that allow the crossovers of the dimension-dependent $\chi$, internal energy~$U$ and  $C_{\rm V}$ in 1D, 2D, and 3D, and of $p$, enthalpy~$H$, heat capacity at constant pressure~$C_{\rm p}$, isothermal bulk modulus $K_T$, and thermal expansion coefficient $\alpha$ in 2D and 3D to be calculated between the low-$T$ degenerate regime and the high-$T$ nondegenerate regime.   Illustratrative plots of the $T$ dependences of these properties are provided.

There are many instances in which free-electron Fermi gas properties are observed in real metals.  An important example is the $T$-dependent $C_{\rm V}$ of metals at low $T$ compared to the Fermi temperature $T_{\rm F}=E_{\rm F}/k_{\rm B}$ ($E_{\rm F}$ is the Fermi energy) which is proportional to $T$ irrespective of the dimensionality of the electron gas in the metal.  However,  the proportionality constant $\gamma$, known as the Sommerfeld electronic heat capacity coefficient, is proportional to the electronic density of states at $E_{\rm F}$ which depends on the dimensionality of the metal for a given electron concentration as discussed below.  The free-electron model as reflected in the theoretical value of $\gamma$ is often fairly close for a particular metal to the measured value and hence is a good starting point for interpreting the $\gamma$ value.  Deviations occur due to features not included in the free-electron Fermi-gas model such as the influences of the interactions of the conduction electrons with the lattice (the electron-phonon interaction) and of electron-electron interactions, such as described in Refs.~\cite{Sevilla2016, Johnston2020}.

In superconductors, the free-electron Fermi-gas model is not appropriate below the superconducting transition temperature $T_{\rm c}$ even for simple metals exhibiting so-called conventional superconductivity because within the BCS theory~\cite{Bardeen1957} the superconductivity arises from indirect attractive interactions between the conduction electrons mediated by phonons.  Here an energy gap opens in the quasiparticle (electron/hole) excitation spectrum below $T_{\rm c}$, so that the electronic heat capacity decreases faster than linearly below $T_{\rm c}$ and approaches zero exponentially for $T\to0$.

Interestingly, the free-electron Fermi-gas theory also sometimes applies to nonmetals containing nuclei with nonzero spin, such as liquid $^3$He for which the nuclei have spin \mbox{$S=1/2$.}  Indeed, as shown in Fig.~\ref{3He_chiFit} below, the magnetic susceptibility $\chi$ versus~$T$ of liquid $^3$He is described well by 3D $S=1/2$ noninteracting-fermion theory all the way from the degenerate ($T\ll T_{\rm F}$) to the nondegenerate ($T\gg T_{\rm F}$) temperature regime, and a fit to the data yielded the value of $T_{\rm F}$ for this material.  Additional examples of 3D Fermi gases include the electron gas in white-dwarf stars and the neutrons in neutron stars~\cite{Kittel2005}, where the Fermi energies were estimated.  Important examples of 2D or quasi-2D electron gases occur in metal-oxide-semiconductor field-effect transistors (MOSFETS) in which the Quantum Hall Effect was discovered~\cite{Klitzing1980} for which the 1985 Nobel Prize in Physics was awarded~\cite{Klitzing1985Nobel}, as well as in other materials~\cite{MOSFETWiki}.

Fundamental expressions and the notation used here are given in the Sec.~\ref{Sec:Fund}.  The equations from which $\mu(T)$ can be calculated for 1D, 2D, and 3D electron Fermi gases are presented in Sec.~\ref{Sec:muvsT}.  The $\chi$ is obtained and plotted versus $T$ from the degenerate to the nondegenerate regime for electron Fermi gases in 1D, 2D, and 3D in Sec.~\ref{Sec:Chi}, and the dimension-dependent $U$ and $C_{\rm V}$ versus~$T$ are calculated and plotted in Sec.~\ref{Sec:Cp}.  The pressure~$p$, enthalpy~$H$, heat capacity at constant pressure~$C_{\rm p}$,  isothermal bulk modulus~$K_T$, and thermal expansion coefficient~$\alpha$ in 2D and 3D are derived and plotted from the degenerate to nondegenerate temperature regime in Secs.~\ref{Sec:Pressure}--\ref{ThermExpand}, respectively.  Concluding remarks are given in Sec.~\ref{ConclRem}.

Some of our results for the free-electron Fermi gas were already known from either published or unpublished sources.  The  $\mu(T)$ in 1D, 2D, and 3D was thoroughly investigated in Ref.~\cite{Cowan2019}.  The $U(T)$ and $C_{\rm V}(T)$ at low~$T$ for 1D, 2D, and 3D were obtained in Ref.~\cite{Cetina1977}.  Sommerfeld expansions of the low-$T$ properties are routinely obtained for all three dimensions in many sources.  Calculations of  $\mu(T)$,  $p(T,V)$, and $C_{\rm V}(T)$  in 1D, 2D, and 3D from the degenerate to the nondegenerate temperature regime are available as of this writing in unpublished lecture notes~\cite{Muller2015}.  Here we include our calculations and plots of these previously-studied quantities for completeness.  The  calculations of $\mu(T)$ in 1D, 2D, and 3D are required for the calculations of all other thermodynamic properties versus dimensionality from the degenerate to the nondegenerate temperature regime.

\section{\label{Sec:Fund} Fundamentals and Notation}

The kinetic energy of an electron of mass $m$ is $E = p^2/2m$, where $p$ is the magnitude of the momentum.  Substituting the quantum-mechanical de Broglie relation $p = \hbar k$ for a matter particle where $k$ is the magnitude of the wave vector, one has
\begin{equation}
E = \frac{\hbar^2 k^2}{2 m},
\end{equation}
which yields the differentials
\begin{eqnarray}
dk &=& \frac{1}{2}\sqrt{\frac{2m}{\hbar^2}}\ \frac{dE}{\sqrt{E}}\nonumber\\*
k\,dk &=& \frac{1}{2}\left(\frac{2m}{\hbar^2}\right)\ dE \label{Eq:Edifferentials} \\*
k^2\,dk &=& \frac{1}{2}\left(\frac{2m}{\hbar^2}\right)^{3/2} \sqrt{E}\ dE.\nonumber
\end{eqnarray}

For traveling waves, using periodic boundary conditions an allowed wave vector has a volume of $2\pi/L$ in 1D, $(2\pi/L)^2 = 4\pi^2/A$ in 2D and $(2\pi/L)^3 = 8 \pi^3/V$ in 3D, where here $A$ is the area and $V$ is the volume of the respective confining box~\cite{Kittel2005}.  Thus a line, circle, or sphere of radius $k$ in {\bf k}-space contains $N_{\rm W}$ traveling waves, given by
\begin{eqnarray}
N_{\rm W} &=& \frac{2k}{2\pi/L}  =\frac{Lk}{\pi}\hspace{0.49in}{\rm (1D)}\nonumber\\*
N_{\rm W} &=& \frac{\pi k^2}{4\pi^2/A}  =\frac{Ak^2}{4\pi}\hspace{0.35in}{\rm (2D)}\label{Eq:NSW2}\\*
N_{\rm W} &=& \frac{(4/3)\pi k^3}{8 \pi^3/V}  =\frac{Vk^3}{6\pi^2}\hspace{0.2in}{\rm (3D)},\nonumber
\end{eqnarray}
where the factor of two in the 1D case arises because $k_x = \pm k$.  When one includes the factor of two Zeeman degeneracy of the free electron, the density of electron states versus energy is $D(E) = 2\,dN_{\rm W}/dE$.  Then Eqs.~(\ref{Eq:NSW2}) yield
\begin{eqnarray}
D(E) &=& \frac{L}{\pi}\sqrt{\frac{2m}{\hbar^2}}\ \frac{1}{\sqrt{E}}   \hspace{0.54in}{\rm (1D)} \nonumber\\*
D(E) &=& \frac{A}{2\pi}\left(\frac{2m}{\hbar^2}\right)  \hspace{0.67in}{\rm (2D)}\label{Eq:D(E)FreeElectrons}\\*
D(E) &=& \frac{V}{2\pi^2}\left(\frac{2m}{\hbar^2}\right)^{3/2} \sqrt{E} \hspace{0.2in}{\rm (3D)}.\nonumber
\end{eqnarray}

We consider an $n$-dimensional volume ($n=1,2,3$) containing $N$ electrons.  At temperature $T = 0$, the $N$ electrons are placed into the energy states $E=\hbar^2k^2/2m$ according to the Pauli Exlusion Principle with the lowest energy states filled first.  Here $m$ is the mass of the free electron.  The energy of the highest-filled state is the Fermi energy $E_{\rm F}$ and the corresponding magnitude of the wave vector is the Fermi wave vector $k_{\rm F}$. The number of filled states $N_{\rm S}$ up to an energy $E$ is given by $N_{\rm S} = \int_0^E D(E)dE$.  Setting $N_{\rm S}$ equal to the number $N$ of electrons and the upper limit of the integral at $T=0$  to $E_{\rm F}$, one obtains
\begin{equation}
N = \int_0^{E_{\rm F}} D(E)dE.
\label{Eq:EF}
\end{equation}
For a given density of states function $D(E)$ one can solve for $E_{\rm F}$ as a function of $N$ after doing the integral.  Also, we have that $E_{\rm F} = \hbar^2 k_{\rm F}^2/(2m)$, so that $k_{\rm F} = \sqrt{2mE/\hbar^2}$.  Substituting the $D(E)$ functions in Eqs.~(\ref{Eq:D(E)FreeElectrons}) into~(\ref{Eq:EF}) and doing the integrals give
\begin{eqnarray}
E_{\rm F} &=& \frac{\hbar^2}{2m}\left(\frac{\pi N}{2L}\right)^2,\hspace{0.2in} k_{\rm F} = \frac{\pi N}{2L}\hspace{0.5in} ({\rm 1D})\nonumber\\*
E_{\rm F} &=& \frac{\hbar^2}{2m}\left(\frac{2\pi N}{A}\right),\hspace{0.2in} k_{\rm F} = \sqrt{\frac{2\pi N}{A}}\hspace{0.3in} ({\rm 2D})\label{EqEFkF}\\*
E_{\rm F} &=& \frac{\hbar^2}{2m}\left(\frac{3\pi^2 N}{V}\right)^{2/3},\hspace{0.2in} k_{\rm F} = \left(\frac{3\pi^2 N}{V}\right)^{1/3},\hspace{0.1in} ({\rm 3D})\nonumber
\end{eqnarray}
where $L$ is the ``volume'' of the system in 1D, $A$ is the ``volume'' in 2D, and $V$ is the volume in 3D\@.  The Fermi velocity (speed) $v_{\rm F}$ is defined as $v_{\rm F} = p_{\rm F}/m = \hbar k_{\rm F}/m$, where $p_{\rm F} = \hbar k_{\rm F}$ is the Fermi momentum of the electron, and hence  $E_{\rm F} = \hbar^2k_{\rm F}^2/2m$.

With the results in Eq.~(\ref{EqEFkF}), one can determine the density of states at the Fermi energy $D(E_{\rm F})$ by substituting the expressions for $E_{\rm F}$ in Eqs.~(\ref{EqEFkF}) for $E$ in Eqs.~(\ref{Eq:D(E)FreeElectrons}), yielding
\begin{eqnarray}
D(E_{\rm F}) &=& \frac{2L}{\pi^2}\left(\frac{2m}{\hbar^2}\right)\left(\frac{L}{N}\right)\hspace{0.55in} ({\rm 1D})\nonumber\\*
D(E_{\rm F}) &=& \frac{A}{2\pi}\left(\frac{2m}{\hbar^2}\right)  \hspace{0.94in}{\rm (2D)}\label{Eq:D(EF)FreeElectrons}\\*
D(E_{\rm F}) &=& \frac{V}{2\pi^2}\left(\frac{2m}{\hbar^2}\right)\left(\frac{3\pi^2N}{V}\right)^{1/3}  \hspace{0.1in}{\rm (3D)}.\nonumber
\end{eqnarray}

Using Eqs.~(\ref{EqEFkF}), one can write the $D(E)$ expressions~(\ref{Eq:D(E)FreeElectrons}) in terms of $E_{\rm F}$ as
\begin{eqnarray}
D(E) &=& \frac{N}{2E_{\rm F}}\sqrt{\frac{E_{\rm F}}{E}}\hspace{0.2in} ({\rm 1D})\nonumber\\*
D(E) &=& \frac{N}{E_{\rm F}}\hspace{0.62in} ({\rm 2D})\label{Eq:D(EF)2}\\*
D(E) &=& \frac{3N}{2E_{\rm F}}\sqrt{\frac{E}{E_{\rm F}}}\hspace{0.2in} ({\rm 3D}).\nonumber
\end{eqnarray}
It is often more convenient to use expressions~(\ref{Eq:D(EF)2}) for $D(E)$ instead of Eqs.~(\ref{Eq:D(E)FreeElectrons}).

The probability that a state at energy $E$ of an electron gas is occupied at temperature $T$ is given by the Fermi distribution function
\begin{equation}
f(E,T)=\frac{1}{e^{(E-\mu)/k_{\rm B}T}+1},
\label{Eq:FDfcn}
\end{equation}
where $\mu$ is the chemical potential that in general depends on $T$\@.  From Eq.~(\ref{Eq:FDfcn}), the chemical potential is the energy at which the probability of the state being occupied by an electron state is 1/2.   One has $\mu(T\to 0) = E_{\rm F}$, the Fermi energy.

In order to simplify notation and facilitate obtaining universal functions and plots, we introduce dimensionless reduced variables
\begin{eqnarray}
\epsilon &\equiv& \frac{E}{E_{\rm F}}\nonumber\\*
\bar{\mu} &\equiv& \frac{\mu}{E_{\rm F}}\label{Eq:DimensionlessUnits}\\*
t &\equiv& \frac{k_{\rm B}T}{E_{\rm F}}.\nonumber
\end{eqnarray}
Then the Fermi function~(\ref{Eq:FDfcn}) becomes
\begin{equation}
f(\epsilon,t)=\frac{1}{e^{(\epsilon-\bar{\mu})/t}+1},
\label{Eq:FDfcn2}
\end{equation}
and the densities of states~(\ref{Eq:D(EF)2}) become
\begin{eqnarray}
D(\epsilon) &=& \frac{N}{2E_{\rm F}}\sqrt{\frac{1}{\epsilon}}\hspace{0.3in} ({\rm 1D})\nonumber\\*
D(\epsilon) &=& \frac{N}{E_{\rm F}}\hspace{0.62in} ({\rm 2D})\label{Eq:D(EF)3}\\*
D(\epsilon) &=& \frac{3N}{2E_{\rm F}}\sqrt{\epsilon} \hspace{0.38in} ({\rm 3D}).\nonumber
\end{eqnarray}
We further define
\[
D(\epsilon)=\frac{N}{E_{\rm F}}{\tt D}(\epsilon),
\]
so the dimensionless reduced density of states ${\tt D}(\epsilon)$ is
\begin{eqnarray}
{\tt D}(\epsilon) &=& \frac{1}{2}\sqrt{\frac{1}{\epsilon}}\hspace{0.34in} ({\rm 1D})\nonumber\\*
{\tt D}(\epsilon) &=& 1\hspace{0.62in} ({\rm 2D})\label{Eq:D(EF)4}\\*
{\tt D}(\epsilon) &=& \frac{3}{2}\sqrt{\epsilon}\hspace{0.42in} ({\rm 3D}).\nonumber
\end{eqnarray}

\section{\label{Sec:muvsT} Temperature Dependence of the Chemical Potential}

\subsection{Overall Temperature Dependence}

In order to calculate the thermodynamic properties of a free-electron Fermi gas over a wide $T$ range from the degenerate to the nondegenerate regimes, one must first determine how the chemical potential~$\mu$  in the Fermi function~(\ref{Eq:FDfcn}) changes with $T$\@.  The chemical potential is determined at each $T$ by the requirement that the number~$N$  of electrons in the Fermi gas be equal to the number of occupied electron states:
\begin{eqnarray}
N &=& \int_0^\infty D(E)f(E,T)dE,\nonumber\\*
{\rm or}\ \ \ 1 &=& \int_0^\infty {\tt D}(\epsilon) f(\epsilon,t)d\epsilon,
\label{Eq:Findmu}
\end{eqnarray}
where $D(E)$ is the dimension-dependent density of states versus energy $E$ including the twofold Zeeman degeneracy of the electron, $\epsilon=E/E_{\rm F}$ is the reduced energy defined in Eq.~(\ref{Eq:DimensionlessUnits}) where the Fermi energy $E_{\rm F}$ is the chemical potential at $T=0$,  ${\tt D}(\epsilon)$ is the reduced density of states in Eq.~(\ref{Eq:D(EF)4}), and $t$ is the reduced temperature $k_{\rm B}T/E_{\rm F}$ where $k_{\rm B}$ is Boltzmann's constant.  The value of $E_{\rm F}$ depends on the linear, areal, and volume density of electrons in 1D, 2D, and 3D, respectively, as given in Eqs.~(\ref{EqEFkF}).

To solve for the reduced chemical potential $\bar{\mu}(t) = \mu(t)/E_{\rm F}$ in Eq.~(\ref{Eq:FDfcn2}) using Eq.~(\ref{Eq:Findmu}), we first find an expression for the definite  integral on the right-hand side of Eq.~(\ref{Eq:Findmu}), which when substituted into Eq.~(\ref{Eq:Findmu}) yields an equation containing $\bar{\mu}$ and $t$ from which $\bar{\mu}(t)$ can be solved for.  Using the ${\tt D}(\epsilon)$ expressions in Eqs.~(\ref{Eq:D(EF)4}), one can evaluate the integral in Eq.~(\ref{Eq:Findmu}) for each of the three dimensions, and Eq.~(\ref{Eq:Findmu}) becomes
\begin{eqnarray}
-\frac{\sqrt{\pi t}}{2}\,{\rm Li}_{1/2}\left(-e^{\bar{\mu}/t}\right)&=& 1\hspace{0.34in} ({\rm 1D})\nonumber\\*
t \ln\left(1+e^{\bar{\mu}/t}\right) &=& 1\hspace{0.35in} ({\rm 2D})\label{Eq:mu(t)}\label{Eq:FindMu}\\*
-\frac{3\sqrt{\pi}\,t^{3/2}}{4}\,{\rm Li}_{3/2}\left(-e^{\bar{\mu}/t}\right) &=& 1\hspace{0.34in} ({\rm 3D}),\nonumber
\end{eqnarray}
where 
\begin{equation}
{\rm Li}_n(x) \equiv \sum_{k=1}^\infty x^k/k^n
\label{Eq:PolyLogFcn}
\end{equation}
is the polylogarithm function~\cite{Mathematica, PolylogWiki}.

According to Eqs.~(\ref{Eq:mu(t)}), for the 2D case $\bar{\mu}(t)$ is given analytically by 
\begin{equation}
\bar{\mu}(t) = t\ln(e^{1/t} - 1)\hspace{0.5in} ({\rm 2D}),
\label{Eq:mubar2D}
\end{equation}
as found previously~\cite{McKelvey1964}.  This immediately gives \mbox{$\bar{\mu}(t\to 0) = 1$} or $\mu(t\to 0) = E_{\rm F}$.  For $t\gg1$ one obtains $\bar{\mu}(t)\approx -t \ln(t)$.  For later use, Eq.~(\ref{Eq:mubar2D}) can be written
\begin{equation}
1+e^{-\bar{\mu}/t} = \frac{1}{1-e^{-1/t}}\hspace{0.37in} ({\rm 2D}).
\label{Eq:expmubart}
\end{equation}
In the 1D and 3D cases, Eqs.~(\ref{Eq:mu(t)}) are solved numerically for $\bar{\mu}(t)$.

\begin{figure}
\includegraphics[width=3.in]{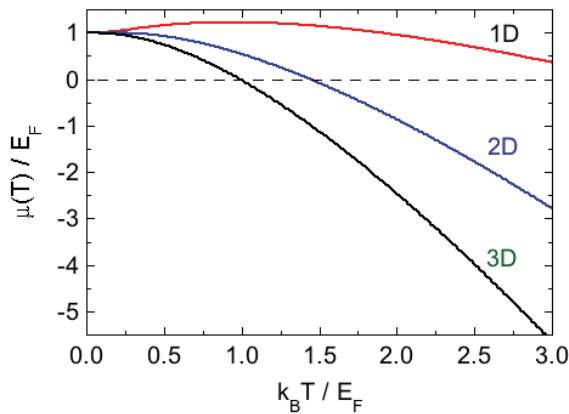}
\caption{Chemical potential $\mu$ divided by the Fermi energy~$E_{\rm F}$ ($\equiv \bar{\mu}$) as a function of  $k_{\rm B}T\equiv t$ for 1D, 2D, and 3D free-electron Fermi gases.  The analytic 2D $T$ dependence is given by Eq.~(\ref{Eq:mubar2D}), and the 1D and 3D $T$ dependences were obtained by numerically solving the respective equation in Eqs.~(\ref{Eq:mu(t)}).}
\label{Fig:Fermi_Gas_mu1D2D3D} 
\end{figure}

Plots of $\bar{\mu}(t)$ for 1D, 2D, and 3D free-electron Fermi gases are shown in Fig.~\ref{Fig:Fermi_Gas_mu1D2D3D}, illustrating the crossover between the degenerate low-$T$ and nondegenerate high-$T$ regime for each dimension.  We note that the low-$T$ Sommerfeld expansion in 2D to any order in $T$  gives the $T$-independent value $\mu=E_{\rm F}$ (Ref.~\cite{Ashcroft1976}, p.~53), contrary to the $T$ dependence in Fig.~\ref{Fig:Fermi_Gas_mu1D2D3D}. 

\subsection{Chemical Potential at High Temperatures}

At temperatures sufficiently high that $\bar{\mu}$ in Fig.~\ref{Fig:Fermi_Gas_mu1D2D3D} becomes strongly negative, the exponential factor $e^{-\bar{\mu}/t}$ in the denominator of the Fermi function~(\ref{Eq:FDfcn2}) becomes much larger than unity, which results in $f(\epsilon,t)\ll 1$ for all electron energies $\epsilon > 0$ of normal physical interest.  One can then ignore the additive factor of unity in the denominator of the Fermi function, which then becomes
\begin{equation}
f(\epsilon,t) = \frac{e^{-\epsilon/t}}{e^{-\bar{\mu}/t}} \equiv \frac{e^{-\epsilon/t}}{Z},
\label{Eq:HighTf}
\end{equation}
which is just the Boltzmann probability  distribution for a classical monatomic ideal gas, where the partition function $Z$ is
\begin{equation}
Z = e^{-\bar{\mu}(t)/t}\gg 1.
\label{Eq:Zgg1}
\end{equation}

Thus $e^{\bar{\mu}(t)/t}\ll 1$, and the Taylor-series expansion~(\ref{Eq:PolyLogFcn}) of the polylogarithm functions in Eqs.~(\ref{Eq:mu(t)}) about $x= 0$ is
\begin{equation}
{\rm Li}_n(-x) = -x+\frac{x^2}{2^n}-\frac{x^3}{3^n}+ \cdots,
\label{Eq:LiExpansion}
\end{equation}
where in Eqs.~(\ref{Eq:mu(t)}) $x = e^{\bar{\mu}/t}\ll1$ for the 1D and 3D cases.  Keeping only the first-order term in $x$, one has ${\rm Li}_n(-x) = -x$ for both $n=1/2$ and~3/2 in 1D and 3D, respectively.  In 2D, we set $e^{1/t}\approx 1 + (1/t)$ in Eq.~(\ref{Eq:mubar2D}).  Thus at high temperatures, Eqs.~(\ref{Eq:mu(t)}) yield
\begin{eqnarray}
\bar{\mu} &=& -\frac{t}{2}\ln\left(\frac{\pi t}{4}\right)\hspace{0.7in}{\rm (1D,\ high}~t)\nonumber\\*
\bar{\mu} &=& -t \ln t\hspace{1.1in}{\rm (2D,\ high}~t)\label{Eq:mubarHighT}\\
\bar{\mu} &=& -t\ln\left(\frac{3\sqrt{\pi}\,t^{3/2}}{4}\right)\hspace{0.38in}{\rm (3D,\ high}~t).\nonumber
\end{eqnarray}

\subsection{Fermi Function}

At low temperatures \mbox{$T\ll T_{\rm F}\equiv E_{\rm F}/k_{\rm B}$} where $T_{\rm F}$ is the Fermi temperature, the Fermi function $f(\epsilon,t)$ is independent of energy except near $E_{\rm F}$ where it decreases rapidly.  This feature suggests that to compute the temperature dependence of physical properties at low temperatures, one should utilize expressions containing $\partial f/\partial \epsilon$ which emphasizes the energy region around $E_{\rm F}$.  Using Eq.~(\ref{Eq:FDfcn2}) for $f(\epsilon,t)$ gives
\begin{eqnarray}
\frac{\partial f(\epsilon,t)}{\partial \epsilon} &=&  -\left(\frac{1}{2t}\right) \frac{1}{1 + \cosh[(\epsilon - \bar{\mu})/t]}.
\label{Eq:dfde}
\end{eqnarray}

The probability that one of the two Zeeman states of the electron within a spatial orbital is occupied by an electron is $f$ and the probability that the other one is unoccupied is $1-f$, so the probability that both events happen at the same time is $f(1-f)$.  Through a straightforward manipulation of $f(\epsilon,t)$, one obtains
\begin{equation}
f(1-f) = -t\frac{\partial f}{\partial \epsilon}.
\label{Eq:f1-f}
\end{equation}

The low-temperature Sommerfeld expansion~\cite{Ashcroft1976, WikiSommerfeld} of the following integral $I(t)$ to order $t^2$ is 
\bse
\begin{equation}
I(t) = \int_0^\infty g(\epsilon)f(\epsilon,t)d\epsilon = \int_0^{\bar{\mu}} g(\epsilon)d\epsilon +  \frac{dg(\epsilon)}{d\epsilon}\Big|_{\epsilon=\bar{\mu}}\frac{\pi^2t^2}{6},
\label{Eq:Int6}
\end{equation}
where here $g(\epsilon<0)=0$ and at low temperatures $\bar{\mu}\approx 1$.  Let $G(\epsilon)$ be the indefinite integral
\bea
G(\epsilon) = \int g(\epsilon)d\epsilon, \ \ {\rm or}\ \  g(\epsilon) = \frac{dG(\epsilon)}{d\epsilon}.
\label{Eq:Geps}
\eea
Then an integration by parts of the first integral in Eq.~(\ref{Eq:Int6}) yields
\bea
I(t) &=& -\int_0^\infty G(\epsilon)\frac{\partial f(\epsilon, t)}{\partial \epsilon} d\epsilon, \nonumber \\
 &=& G(\bar{\mu}) + \frac{d^2G(\epsilon)}{d\epsilon^2}\bigg{|}_{\epsilon = \bar{\mu}}\frac{\pi^2t^2}{6}.\label{Eq:Int22}
\eea
\ese
Equations~(\ref{Eq:Int6}) and~(\ref{Eq:Int22}) are useful in different contexts. 

\subsection{Chemical Potential at Low Temperatures}

The chemical potential is determined by Eq.~(\ref{Eq:Findmu}).  To evaluate the integral in Eq.~(\ref{Eq:Findmu}) in terms of the low-temperature expansion in Eq.~(\ref{Eq:Int6}), the function $g(\epsilon) = {\tt D}(\epsilon)$, where ${\tt D}(\epsilon)$ is given in Eqs.~(\ref{Eq:D(EF)4}).  Thus we have
\begin{eqnarray}
\int_0^{\bar{\mu}}{\tt D}(\epsilon)d\epsilon &=& \sqrt{\bar{\mu}},\ \ \ \ \frac{d{\tt D}(\epsilon)}{d\epsilon} = -\frac{1}{4\bar{\mu}^{3/2}}\hspace{0.05in} ({\rm 1D})\nonumber\\*
\int_0^{\bar{\mu}}{\tt D}(\epsilon)d\epsilon &=& \bar{\mu},\ \ \ \ \ \ \ \frac{d{\tt D}(\epsilon)}{d\epsilon} = 0\hspace{0.42in} ({\rm 2D})\label{Eq:IntD(EF)}\\*
\int_0^{\bar{\mu}}{\tt D}(\epsilon)d\epsilon &=& \bar{\mu}^{3/2},\ \ \ \ \frac{d{\tt D}(\epsilon)}{d\epsilon} = \frac{3}{4\sqrt{\bar{\mu}}}\hspace{0.15in} ({\rm 3D}).\nonumber
\end{eqnarray}
Inserting these results into Eq.~(\ref{Eq:Int6}) gives
\begin{eqnarray}
\int_0^{\bar{\mu}}{\tt D}(\epsilon)f(\epsilon,t)d\epsilon &=& \sqrt{\bar{\mu}}-\frac{\pi^2t^2}{24\bar{\mu}^{3/2}}\hspace{0.05in} ({\rm 1D})\nonumber\\*
\int_0^{\bar{\mu}}{\tt D}(\epsilon)f(\epsilon,t)d\epsilon &=& \bar{\mu}\hspace{0.78in} ({\rm 2D})\label{Eq:IntD(EF)2}\\*
\int_0^{\bar{\mu}}{\tt D}(\epsilon)f(\epsilon,t)d\epsilon &=& \bar{\mu}^{3/2}+ \frac{\pi^2t^2}{8\sqrt{\bar{\mu}}}\hspace{0.15in} ({\rm 3D}).\nonumber
\end{eqnarray}

\subsubsection{One-Dimensional Fermi Gas}

Setting the integral for 1D in Eq.~(\ref{Eq:IntD(EF)2}) equal to unity according to Eq.~(\ref{Eq:Findmu}) gives
\begin{equation}
\bar{\mu}^2-\bar{\mu}^{3/2}=\frac{\pi^2t^2}{24}.
\label{Eq:mut21D}
\end{equation}
This is a 4$^{\rm th}$-order equation in $\bar{\mu}^{1/2}$.  However, we take advantange of the fact that $\bar{\mu}$ approaches unity at $t\to 0$.  Thus we write 
\begin{equation}
\bar{\mu}(t)=1+\alpha(t)
\label{Eq:Defalpha}
\end{equation}
with $|\alpha(t)| \ll 1$ and use the Taylor-series expansion 
\[
(1+\alpha)^n \approx 1+n\alpha 
\]
to obtain from Eq.~(\ref{Eq:mut21D})
\[
(1+2\alpha) - \left(1+\frac{3}{2}\alpha\right) = \frac{\alpha}{2}=\frac{\pi^2t^2}{24}.
\]
Substituting $\alpha = \bar{\mu}-1$ gives
\begin{equation}
\bar{\mu} = 1 + \frac{\pi^2t^2}{12} \hspace{0.2in} ({\rm 1D}).
\label{Eq:mu1DLoT}
\end{equation}
This increase in $\bar{\mu}$ with increasing $t$ at small $t$ agrees with the 1D plot in Fig.~\ref{Fig:Fermi_Gas_mu1D2D3D}.

\subsubsection{Two-Dimensional Fermi Gas}

According to Eq.~(\ref{Eq:mubar2D}), for $t\ll1$ one just obtains
\bea
\bar{\mu}=1,
\label{Eq:mubar2DLowT}
\eea
indicating that there is no power-law Sommerfeld expansion for $\bar{\mu}(T)$ at low~$T$, in agreement with Ref.~\cite{Ashcroft1976}, p.~53.

\subsubsection{Three-Dimensional Fermi Gas}

Setting the integral for 3D in Eq.~(\ref{Eq:IntD(EF)2}) equal to unity according to Eq.~(\ref{Eq:Findmu}) gives
\begin{equation}
\bar{\mu}^2-\bar{\mu}^{1/2}=-\frac{\pi^2t^2}{8}.
\label{Eq:mut21D2}
\end{equation}
Then following the same steps as for the 1D case gives
\begin{equation}
\bar{\mu} = 1 - \frac{\pi^2t^2}{12} \hspace{0.2in} ({\rm 3D}).
\label{Eq:mu3DLoT}
\end{equation}
This decrease in $\bar{\mu}$ with increasing $t$ is in agreement with the 3D plot in Fig.~\ref{Fig:Fermi_Gas_mu1D2D3D} for small $t$.  Interestingly, the temperature dependence of $\bar{\mu}$ for 3D has the same magnitude but opposite sign as that in Eq.~(\ref{Eq:mu1DLoT}) for 1D, and hence they are symmetrically disposed with respect to the (nearly) temperature-independent behavior at low $t$ for the 2D case.

\section{\label{Sec:Chi} Magnetic Spin Susceptibility}

For a single free electron with spin $S = 1/2$, the magnetic spin susceptibility follows the Curie law
\bse
\bea
\chi_1 = \frac{C_1}{T}.
\label{Eq:ChiOneS122}
\eea
The single-spin Curie constant $C_1$ for spin $S=1/2$ is
\bea
C_1=\frac{g^2S(S+1)\mu_{\rm B}^2}{3k_{\rm B}}=\frac{g^2\mu_{\rm B}^2}{4k_{\rm B}},
\label{Eq:C1}
\eea
\ese
where $g\approx2$ is the spectroscopic splitting factor of the electron magnetic moment and $\mu_{\rm B}$ is the Bohr magneton.  Here, with respect to its spin, the term ``free electron'' means that the $z$-component of its magnetic moment is free to fluctuate between its two spin magnetic quantum numbers $m_S=\pm1/2$.  Such a spin is an ``unpaired'' spin.  In a free-electron gas, a spin is unpaired if another electron does not occupy the same orbital with the other $m_S$ state, because according to the Pauli Exclusion Principle the electron is then free to fluctuate back and forth between its two $\mu_z = -gm_S\mu_{\rm B}$ $z$-component magnetic-moment values.  The probability that an electron occupies a spatial orbital with a particular spin state ($\mu_z$ value) at energy $E$ and temperature $T$ is given by the Fermi function $f(E,T)$.  The probability that the other $\mu_z$ state is not occupied is $1-f(E,T)$.  Thus the probability $P_{\rm unpaired}(E,T)$ that an electron is unpaired  is given by 
\[
P_{\rm unpaired}(E,T) = f(E,T)\, [1-f(E,T)].
\]

The  magnetic susceptibility of the electron gas is then
\begin{equation}
\chi=N_{\rm unpaired}\frac{C_1}{T} = N_{\rm unpaired}\frac{g^2\mu_{\rm B}^2}{4k_{\rm B}T}.
\label{Eq:Chiegas}
\end{equation}
In dimensionless reduced temperature units $t=k_{\rm B}T/E_{\rm F}$ given in Eqs.~(\ref{Eq:DimensionlessUnits}), Eq.~(\ref{Eq:Chiegas}) becomes
\begin{equation}
\chi=N_{\rm unpaired}\frac{g^2\mu_{\rm B}^2}{4E_{\rm F}t}.
\label{Eq:Chiegas2}
\end{equation}
The number of unpaired electrons at temperature~$T$ is
\begin{eqnarray}
N_{\rm unpaired}(T) &=& \int_0^\infty P_{\rm unpaired}(E,T)D(E)dE \nonumber\\*
&=& \int_0^\infty f(E,T)[1-f(E,T)]D(E)dE.\nonumber
\end{eqnarray}
In dimensionless variables one has
\[
\frac{N_{\rm unpaired}(t)}{N}= \int_0^\infty f(\epsilon,t)[1-f(\epsilon,t)]{\tt D}(\epsilon)\,d\epsilon.
\]
Then Eq.~(\ref{Eq:f1-f}) yields
\bea
\frac{N_{\rm unpaired}(t)}{N} &=& -t\int_0^\infty {\tt D}(\epsilon)\frac{\partial f(\epsilon,t)}{\partial\epsilon}\,d\epsilon \label{Eq:Nunpaired}\\
&=& \frac{1}{2} \int_0^\infty \frac{{\tt D}(\epsilon)}{1+\cosh[(\epsilon-\bar{\mu})/t]}d\epsilon, \nonumber
\eea
where Eq.~(\ref{Eq:dfde}) was used to obtain the second equality.

For $t\ll1$, we write Eq.~(\ref{Eq:Nunpaired}) as
\begin{eqnarray}
\lim_{t\to0}\frac{N_{\rm unpaired}(t)}{N} &=& t\int_0^\infty {\tt D}(\epsilon)\,\delta(\epsilon-1)\,d\epsilon\nonumber\\*
 &=& t\,{\tt D}(\epsilon=1),
\label{Eq:Nunpaired2}
\end{eqnarray}
where $\delta(x)$ is the Dirac delta function.  Switching back to conventional quantities using the conversion expressions in Sec.~\ref{Sec:Fund} gives
\[
\lim_{T\to0}N_{\rm unpaired}= k_{\rm B}T\, D(E_{\rm F}).
\]
Then using Eq.~(\ref{Eq:Chiegas}) one obtains the well-known result for the Pauli spin susceptibility
\begin{eqnarray}
\chi^{\rm Pauli} &\equiv& \chi(t=0) = [k_{\rm B}T D(E_{\rm F})]\frac{g^2\mu_{\rm B}^2}{4k_{\rm B}T}\nonumber\\*
&=& \frac{g^2\mu_{\rm B}^2}{4}D(E_{\rm F}) =  \frac{Ng^2\mu_{\rm B}^2}{4E_{\rm F}}{\tt D}(1).\label{Eq:ChiPauli}
\end{eqnarray}
Then Eqs.~(\ref{Eq:D(EF)4}) yield
\bse
\label{Eqs:chi01D2D3D}
\bea
\chi(0) &=& \frac{Ng^2\mu_{\rm B}^2}{8E_{\rm F}}\quad ({\rm 1D})\\
 &=& \frac{Ng^2\mu_{\rm B}^2}{4E_{\rm F}}\quad ({\rm 2D})\\
 &=& \frac{3Ng^2\mu_{\rm B}^2}{8E_{\rm F}}\quad ({\rm 3D}).
\eea
\ese

\begin{figure}
\includegraphics[width=3.in]{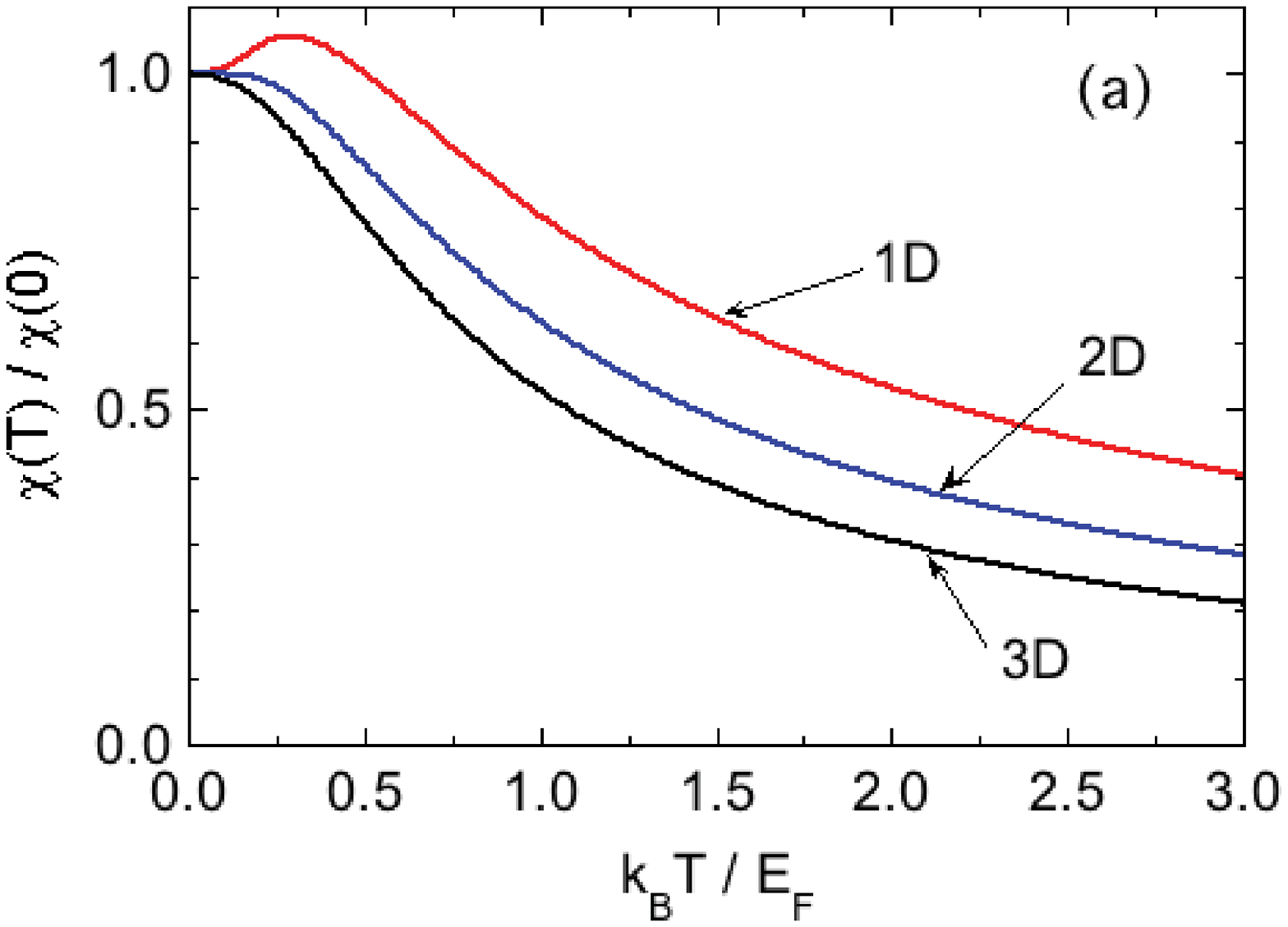}
\includegraphics[width=3.in]{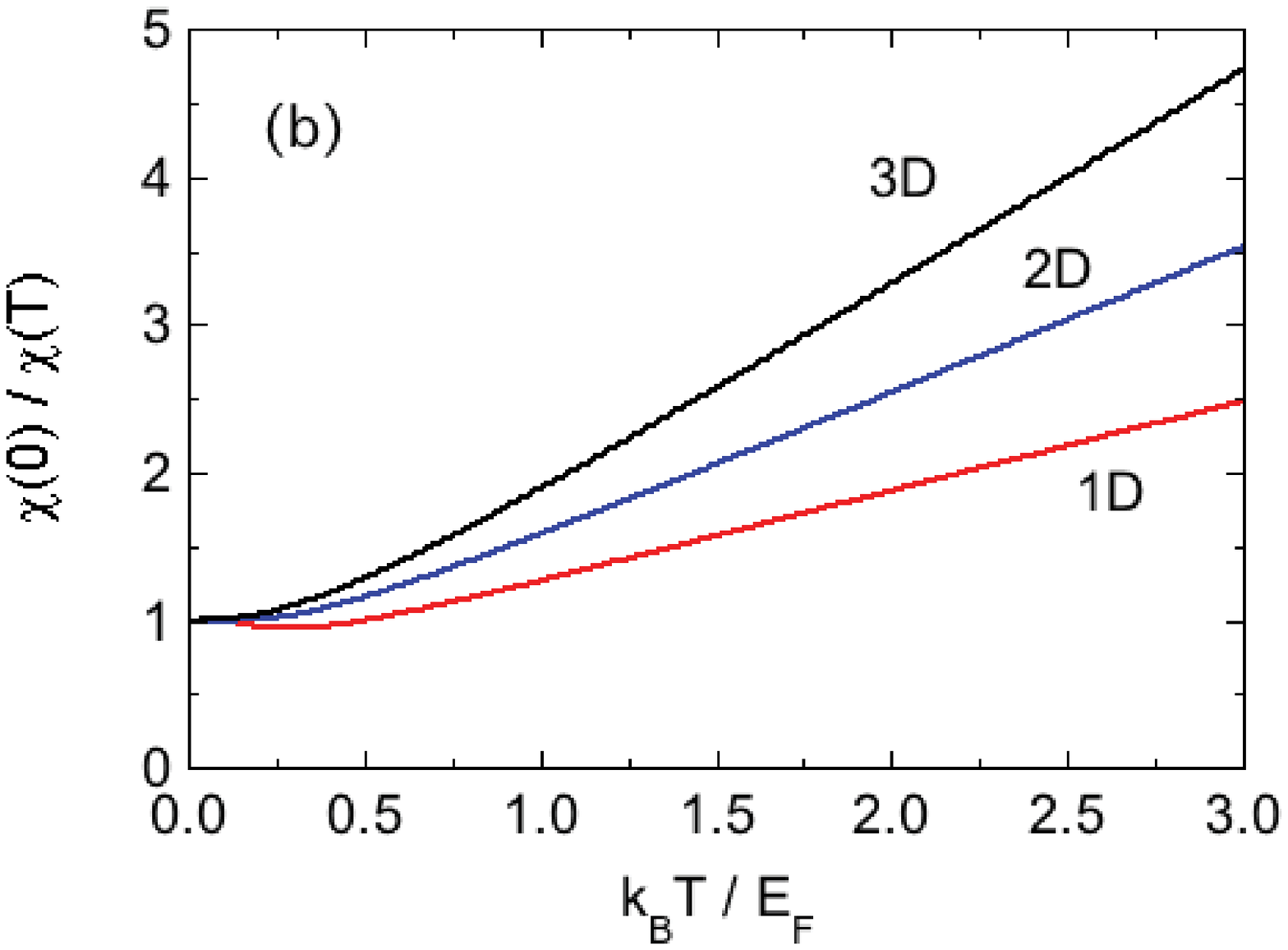}
\caption{(a)~Normalized magnetic spin susceptibility $\chi(t)/\chi(0)$ versus reduced temperature $t = k_{\rm B}T/E_{\rm F}$ for $S=1/2$ Fermi gases in 1D, 2D and 3D, where $\chi(0)$ is the respective zero-temperature susceptibility.  (b)~Inverse normalized susceptibility versus $t$.  At high finite temperatures, the susceptibilities in 1D and 3D approach a Curie-law  behavior $C/T$ whereas the data in 2D approach Curie-Weiss behavior $C/(T-\theta_{\rm 2D})$.  The negative $\theta_{\rm 2D}$ value arises from the Fermi statistics rather than from electron interactions.}
\label{Fig:Fermi_Gas_1D2D3D_chi}
\end{figure}

\begin{figure}
\includegraphics[width=3.3in]{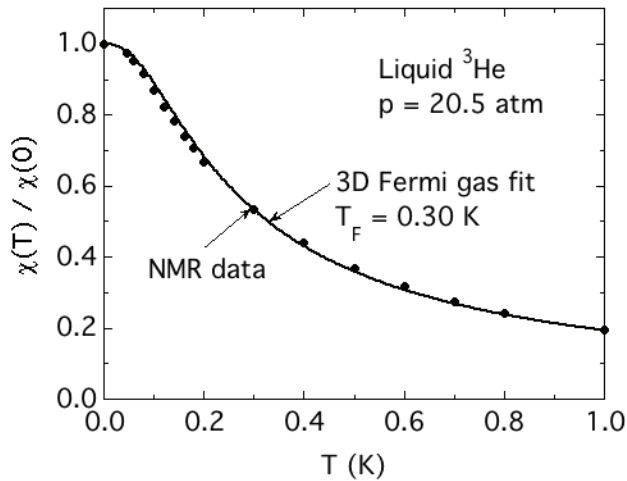}
\caption{Nuclear magnetic susceptibility data normalized by the value at $T\to0$, $\chi(T)/\chi(0)$, obtained from NMR measurements versus temperature~$T$ for liquid $^3$He under a pressure of 20.5 atm.\ (filled circles), where the data point at $T=0$ is extrapolated~\cite{Thomson1962}.  A fit of the data by Eq.~(\ref{Eq:chit0})  is shown by the solid curve where the fitted Fermi temperature is 0.30~K\@.}
\label{3He_chiFit} 
\end{figure}

Using Eqs.~(\ref{Eq:Chiegas}), (\ref{Eq:Nunpaired}), and (\ref{Eq:ChiPauli}), the normalized dimension-dependent susceptibility in reduced variables is given by
\bea
\frac{\chi(t)}{\chi(0)} = \frac{1}{2t\,{\tt D}(1)}\int_0^\infty \frac{{\tt D}(\epsilon)}{1+\cosh[(\epsilon-\bar{\mu})/t]}d\epsilon,
\label{Eq:chit0}
\eea
where the dimension-dependent reduced densities of states ${\tt D}(\epsilon)$ are given in Eq.~(\ref{Eq:D(EF)4}).  Figure~\ref{Fig:Fermi_Gas_1D2D3D_chi}(a) shows plots of $\chi(t)/\chi(0)$ for 1D, 2D, and 3D $S=1/2$ Fermi gases from $t=0$ to~3.  The dimensionality of the Fermi gas is seen to significantly affect the results.  The inverse normalized susceptibilities are shown in Fig.~\ref{Fig:Fermi_Gas_1D2D3D_chi}(b) which will be discussed in Sec.~\ref{C WLaw}.

Figure~\ref{3He_chiFit}  shows our fit to the measured~\cite{Thomson1962} spin susceptibility of the $S=1/2$ nuclei in liquid $^3$He under a pressure of 20.5~atm which illustrates the crossover from the degenerate regime at 0.045~K to the nondegenerate regime at 1.0~K\@.  The single-parameter fit yielded the Fermi temperature $T_{\rm F} = 0.30$~K\@.

In an applied magnetic field~$H$, the Zeeman energy levels of the $^3$He nuclei with azimuthal quantum number $m_S=\pm 1/2$ are split by the $H$ according to $\Delta E = \gamma H$, where $\gamma$ is the nuclear gyromagnetic ratio.  For the fixed angular frequency $\omega$ of the rf magnetic field applied to the sample, resonance occurs when the quantum of energy of the field $\hbar \omega$ equals $\Delta E$ ($\hbar$ is Planck's constant divided by 2$\pi$), at which frequency power is absorbed from the rf source. Plotting the absorbed power versus $H$ at a given $T$ yields the rf absorption line which has a certain linewidth.  The integral of the absorption line versus~$H$ is proportional to the magnetic susceptibility of the sample at the given $T$\@.  Then normalizing the area of the absorption line to that in the ``high''-$T$ Curie-law region at 1~K yields the $\chi(T)/\chi(T=0)$ data plotted in Fig.~\ref{3He_chiFit}.

\subsection{Low-Temperature Behavior}

In order to obtain the lowest-order power-law temperature dependence of the magnetic susceptibility at low temperatures in 1D and 3D, we note that the integral in Eq.~(\ref{Eq:Nunpaired}) has the form of the Sommerfeld expansion in Eq.~(\ref{Eq:Int22}) with $G(\epsilon) = {\tt D}(\epsilon)$.  Therefore in 1D and 3D we use Eqs.~(\ref{Eq:Int22}) and~(\ref{Eq:Nunpaired}) to obtain 
\bea
\frac{N_{\rm unpaired}(t)}{N}= t\left[{\tt D}(\bar{\mu}) + \frac{d^2 {\tt D}(\epsilon)}{d \epsilon^2}\Big|_{\bar{\mu}}\frac{\pi^2t^2}{6}\right].
\label{Eq:Nunp5}
\eea
Then from Eq.~(\ref{Eq:Chiegas2}) the spin susceptibility is
\begin{equation}
\chi=\frac{Ng^2\mu_{\rm B}^2}{4E_{\rm F}}\left[{\tt D}(\bar{\mu}) + \frac{d^2 {\tt D}(\epsilon)}{d \epsilon^2}\Big|_{\bar{\mu}}\frac{\pi^2t^2}{6}\right].
\label{Eq:Chiegas3}
\end{equation}
In 2D we obtain an analytic result for all~$T$ below.

\subsubsection{1D Electron Gas} 

In one dimension, from Eqs.~(\ref{Eq:D(EF)4}) we have
\begin{equation}
{\tt D}(\bar{\mu}) = \frac{1}{2}\bar{\mu}^{-1/2}
\label{Eq:D(mu)}
\end{equation}
and
\bea
\frac{d^2 {\tt D}(\epsilon)}{d \epsilon^2}\Big|_{\bar{\mu}}= \frac{3}{8}\bar{\mu}^{-5/2}.
\eea
The 1D $t$ dependence of $\bar{\mu}$ at low temperatures is given by Eq.~(\ref{Eq:mu1DLoT}).  Then using the Taylor-series expansion \mbox{$(1+x)^n \approx 1+nx$} gives, to order $t^2$,
\bse
\begin{eqnarray}
{\tt D}(\bar{\mu}) &=& \frac{1}{2}\left( 1 - \frac{\pi^2t^2}{24}\right).\\*
\frac{d^2 {\tt D}(\epsilon)}{d \epsilon^2}\Big|_{\bar{\mu}} t^2 &=& \frac{3}{8}t^2.
\end{eqnarray}
\ese
Inserting these results into Eq.~(\ref{Eq:Chiegas3}) using Eqs.~(\ref{Eq:ChiPauli}) and~(\ref{Eq:D(EF)4}) gives
\begin{equation}
\chi(t)=\chi_{\rm 1D}(0)\left(1 + \frac{\pi^2t^2}{12}\right).
\label{Eq:Chiegas5}
\end{equation}

\subsubsection{3D Electron Gas}

In three dimensions, from Eqs.~(\ref{Eq:D(EF)4}) one has
\begin{equation}
{\tt D}(\bar{\mu}) = \frac{3}{2}\sqrt{\bar{\mu}}.
\label{Eq:D(mu)2}
\end{equation}
We then obtain
\[
\frac{d^2 {\tt D}(\epsilon)}{d \epsilon^2}\Big|_{\bar{\mu}}=-\frac{3}{8}\bar{\mu}^{-3/2}.
\]
The reduced 3D chemical potential at low temperatures is given by Eq.~(\ref{Eq:mu3DLoT}).  Again using the Taylor series expansion $(1+x)^n \approx 1 + nx$ for $x\ll1$ gives, to order $t^2$,
\begin{eqnarray}
{\tt D}(\bar{\mu}) &=& \frac{3}{2}\left( 1 - \frac{\pi^2t^2}{24}\right).\\*
\frac{d^2 {\tt D}(\epsilon)}{d \epsilon^2}\Big|_{\bar{\mu}}t^2 &=& -\frac{3}{8}t^2.
\end{eqnarray}
Inserting these results into Eq.~(\ref{Eq:Chiegas3}) using Eqs.~(\ref{Eq:ChiPauli}) and~(\ref{Eq:D(EF)4}) gives, to order $t^2$,
\begin{equation}
\chi(t)=\chi_{\rm 3D}(0)\left(1- \frac{\pi^2t^2}{12}\right).
\label{Eq:Chiegas7}
\end{equation}

Thus the 1D and 3D temperature behaviors in Eqs.~(\ref{Eq:Chiegas5}) and~(\ref{Eq:Chiegas7}), respectively, have the same temperature dependence magnitude but with opposite signs at low temperatures.  These behaviors agree with the respective plots in Fig.~\ref{Fig:Fermi_Gas_1D2D3D_chi}(a) at low temperatures.

\subsubsection{\label{Sec:2DEG} 2D Electron Gas}

For an electron gas in two dimensions an analytic expression for the spin susceptibility can be obtained.  The 2D case is special because the density of states is independent of the energy of the $S=1/2$ fermion, as shown in Eqs.~(\ref{Eq:D(E)FreeElectrons}) and~(\ref{Eq:D(EF)4}).  Therefore Eq.~(\ref{Eq:Nunpaired}) becomes
\begin{eqnarray}
\frac{N_{\rm unpaired}(T)}{N} &=& -t{\tt D}(\epsilon)\int_0^\infty df(\epsilon) \nonumber\\*
&=& -t{\tt D}(\epsilon)[f(\epsilon=\infty)-f(\epsilon=0)].\nonumber\\
\label{Eq:Nunpaired2D}
\end{eqnarray}
Since $f(\epsilon,t)=1/[e^{(\epsilon - \bar{\mu})/t}+1]$, one has
\[
f(\epsilon=\infty)= 0
\]
and
\[
f(\epsilon=0)= \frac{1}{e^{-\bar{\mu}/t}+1} = 1-e^{-1/t},
\]
where the latter equality was given in Eq.~(\ref{Eq:expmubart}).  Then Eq.~(\ref{Eq:Nunpaired2D}) becomes
\begin{equation}
\frac{N_{\rm unpaired}(T)}{N} = t{\tt D}(\epsilon)\left(1-e^{-1/t}\right).
\label{Eq:Nunpaired2D2}
\end{equation}
Using ${\tt D}(\epsilon) = 1$ and $t=k_{\rm B}T/E_{\rm F}$ from Sec.~\ref{Sec:Fund} gives
\begin{equation}
N_{\rm unpaired}(T) = N\frac{k_{\rm B}T}{E_{\rm F}}\left(1-e^{-E_{\rm F}/k_{\rm B}T}\right).
\label{Eq:Nunpaired2D3}
\end{equation}
Finally, substituting this into Eq.~(\ref{Eq:Chiegas}) gives
\begin{subequations}
\label{Eqs:chi2DT}
\begin{equation}
\chi = \frac{Ng^2\mu_{\rm B}^2}{4E_{\rm F}}\left(1-e^{-E_{\rm F}/k_{\rm B}T}\right),
\label{Eq:chi2Dfull}
\end{equation}
where using Eqs.~(\ref{Eq:D(EF)3}) one obtains
\begin{equation}
\chi(T=0) = \frac{Ng^2\mu_{\rm B}^2}{4E_{\rm F}} = \frac{g^2}{4}\mu_{\rm B}^2D(E_{\rm F}),
\end{equation}
\end{subequations}
as in the general Eq.~(\ref{Eq:ChiPauli}).  The $\chi(t)/\chi(0)$ from Eqs.~(\ref{Eqs:chi2DT}) is plotted versus $t$ in Fig.~\ref{Fig:Fermi_Gas_1D2D3D_chi}(a) along with the 1D and 3D results.

\begin{figure}
\includegraphics[width=3.3in]{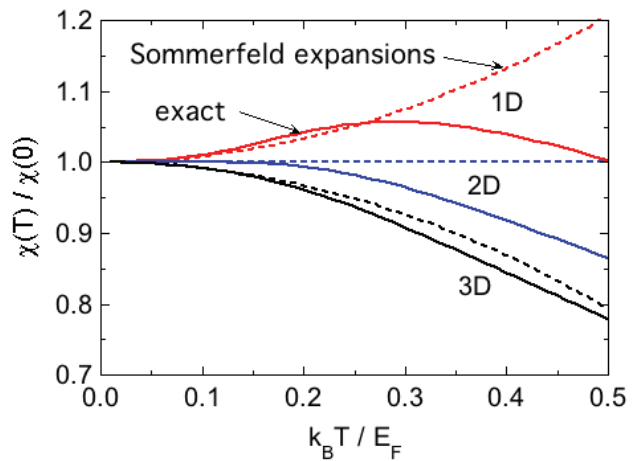}
\caption{Comparison of the numerically-exact calculations of the normalized magnetic susceptibility $\chi(T)/\chi(0)$ in 1D, 2D, and 3D (solid curves) with the predictions of Sommerfeld expansions (dashed curves) at low temperatures relative to the Fermi temperature~$E_{\rm F}/k_{\rm B}$.}
\label{Fermi_Gas_1D2D3D_chi_LowT} 
\end{figure}

A Sommerfeld expansion of the low-$T$ magnetic susceptibility behavior in 2D to any order in $t$ gives $\chi(t)/\chi(0)=1$~\cite{Ashcroft1976}.  Comparisons of the numerically-exact predictions of the normalized susceptibilities $\chi(t)/\chi(0)$ in 1D, 2D, and 3D in Fig.~\ref{Fig:Fermi_Gas_1D2D3D_chi} at low temperatures with the Sommerfeld expansion predictions are shown in Fig.~\ref{Fermi_Gas_1D2D3D_chi_LowT}.  The agreement of these expansions with the exact results is seen to be limited to the temperature range $t\lesssim 0.1$.

\subsection{\label{C WLaw} High-Temperature Behavior}

According to Eq.~(\ref{Eq:HighTf}), at very high temperatures the Fermi function becomes the Boltzmann distribution
\[
f(\epsilon,t) = \frac{e^{-\epsilon/t}}{Z},
\]
where $Z(t) = e^{-\bar{\mu}/t}$ is the partition function.  Then
\begin{equation}
\frac{\partial f(\epsilon,t)}{\partial\epsilon} = -\frac{1}{t}\,\frac{e^{-\epsilon/t}}{Z} = -\frac{1}{t}f(\epsilon,t),
\end{equation}
and using Eq.~(\ref{Eq:Findmu}), Eq.~(\ref{Eq:Nunpaired}) becomes 
\begin{equation}
\frac{N_{\rm unpaired}(t)}{N}= \int_0^\infty {\tt D}(\epsilon)f(\epsilon,t)\,d\epsilon = 1.
\label{Eq:NunpairedHighT}
\end{equation}
Thus for $k_{\rm B}T\gg E_{\rm F}$, the number of unpaired electrons is equal to the total number $N$ of electrons in the system, and the spin susceptibility is just that of $N$ free spins-1/2, 
\bea
\chi = \frac{NC_1}{T} = \frac{Ng^2\mu_{\rm B}^2}{4k_{\rm B}T} = \frac{Ng^2\mu_{\rm B}^2}{4E_{\rm F}t},
\label{Eq:chiNC1T}
\eea
which is the same for 1D, 2D, and 3D free-electron gases, where $C_1$ is the single-electron Curie constant in Eq.~(\ref{Eq:C1}).

It is of interest to calculate the spin susceptibility in 2D at high temperatures, but not in the limit $T\to\infty$, by keeping one additional term in the Taylor-series expansion of the exponential, i.e., $e^x = 1 + x + x^2/2$.  Then Eq.~(\ref{Eq:chi2Dfull}) becomes
\begin{eqnarray}
\chi &=& \frac{Ng^2\mu_{\rm B}^2}{4E_{\rm F}}\left[\frac{E_{\rm F}}{k_{\rm B}T}- \frac{1}{2}\left(\frac{E_{\rm F}}{k_{\rm B}T}\right)^2\right]\label{Eq:chi2DhiT}\\*
&=& \frac{Ng^2\mu_{\rm B}^2}{4k_{\rm B}T}\left(1-\frac{1}{2}\frac{E_{\rm F}}{k_{\rm B}T}\right).\nonumber
\end{eqnarray}
Using the Taylor-series expansion $(1+x)^n \approx 1+nx$ for small $x$ which gives $1-x \approx 1/(1+x)$ with $x = E_{\rm F}/(2k_{\rm B}T$), Eq.~(\ref{Eq:chi2DhiT}) can be written
\begin{eqnarray}
\chi &=& \frac{Ng^2\mu_{\rm B}^2}{4k_{\rm B}T\left(1+\frac{E_{\rm F}}{2k_{\rm B}T}\right)} = \frac{Ng^2\mu_{\rm B}^2}{4k_{\rm B}\left(T+\frac{E_{\rm F}}{2k_{\rm B}}\right)}.
\end{eqnarray}
This has the form of the Curie-Weiss law 
\begin{equation}
\chi=\frac{C}{T-\theta_{\rm 2D}},
\label{Eq:chi2DhighT}
\end{equation}
where $C=Ng^2\mu_{\rm B}^2/(4k_{\rm B})$ is the Curie constant of the $S=1/2$ Fermi gas and the Weiss temperature is
\bea
\theta_{\rm 2D} = -E_{\rm F}/2k_{\rm B}.
\eea
In reduced units one has
\bea
\bar{\theta}_{\rm 2D} \equiv \frac{k_{\rm B}\theta_{\rm 2D}}{E_{\rm F}} = -\frac{1}{2}.
\label{Ea:Theta2D}
\eea
The negative value agrees with expectation from Fig.~\ref{Fig:Fermi_Gas_1D2D3D_chi}(b) obtained by extrapolating the nearly linear $\chi(0)/\chi(t)$ data for $t > 1$ to the horizontal axis.  In contrast to the usual interpretation of the Weiss temperature, the Weiss temperature  does not arise from interactions between the spins but rather from the temperature-dependent Fermi statistics. 

\begin{figure}
\includegraphics[width=3.in]{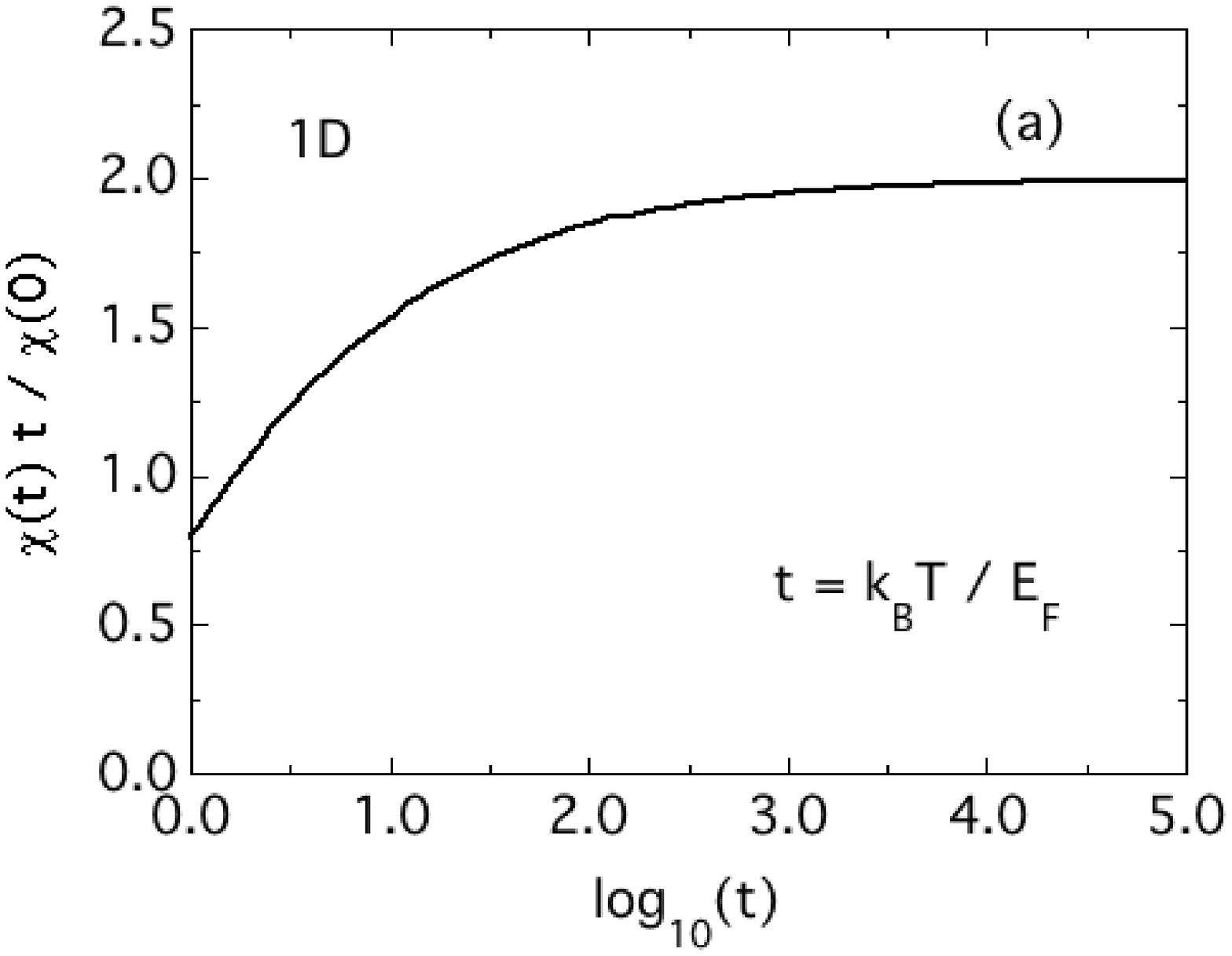}
\includegraphics[width=3.in]{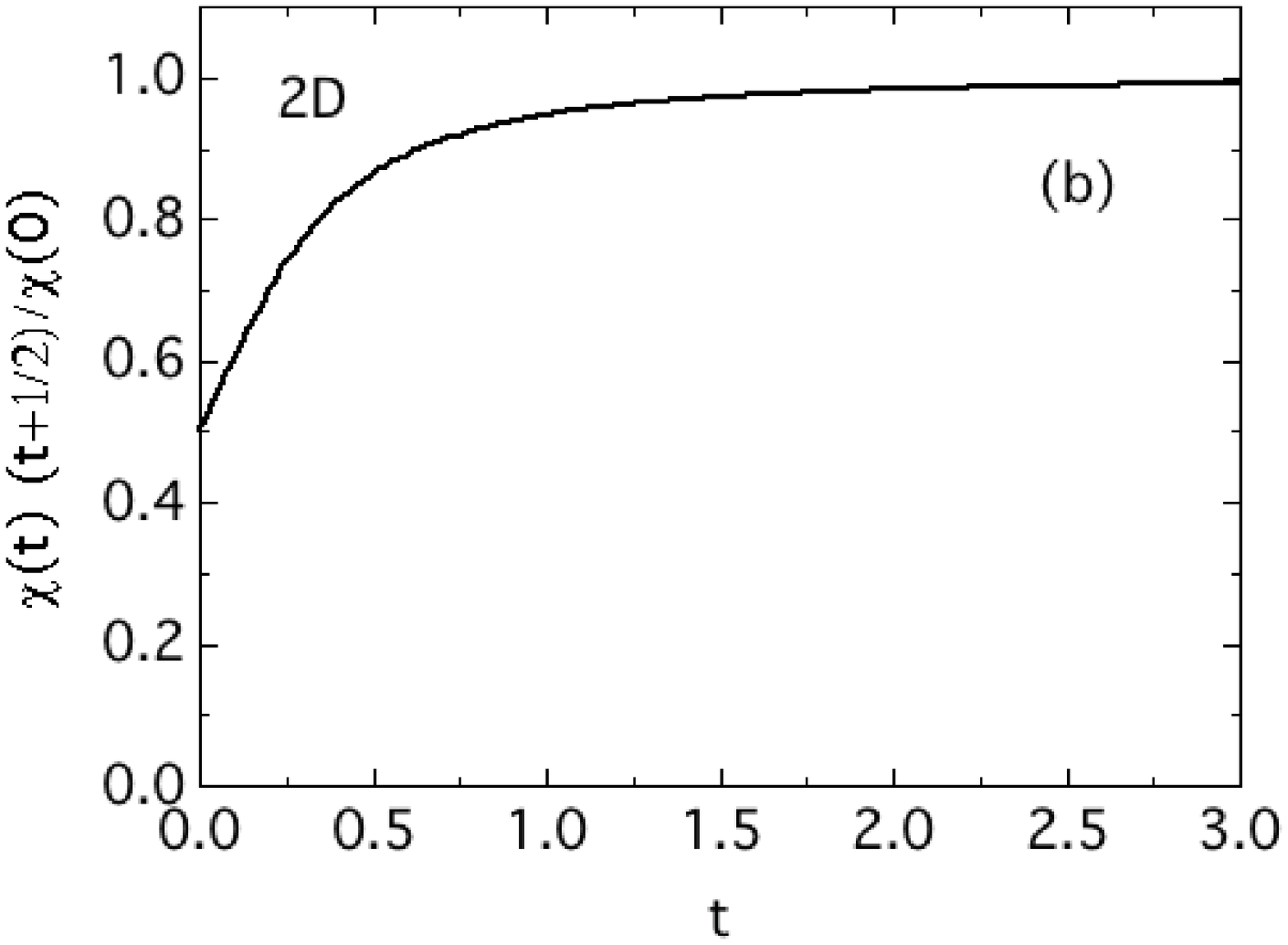}
\includegraphics[width=3.in]{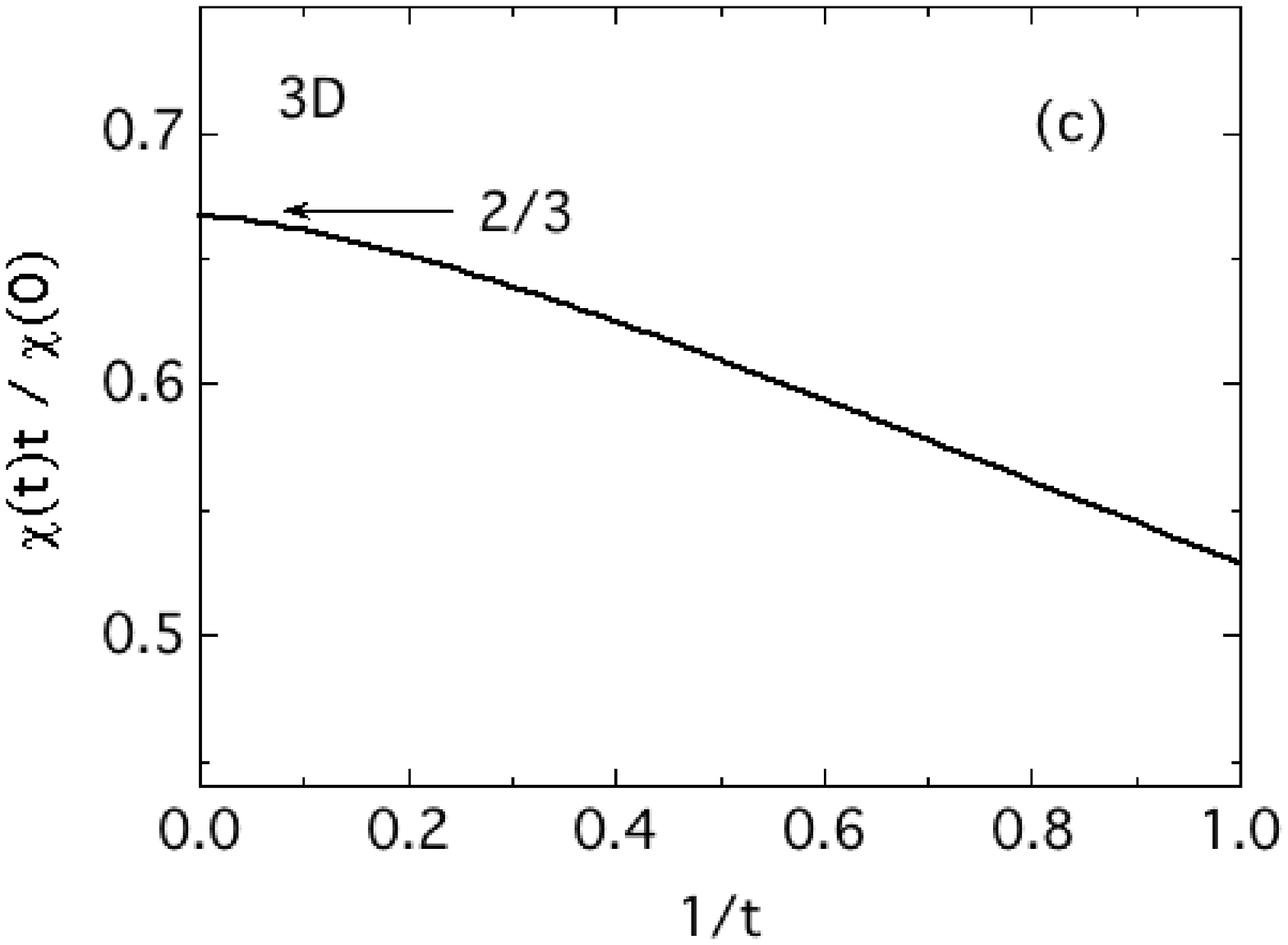}
\caption{Approach of the normalized susceptibilities $\chi(t) t/\chi(0)$ in 1D and 3D and $\chi(t)(t-\bar{\theta}_{\rm 2D})/\chi(0)$ in 2D to the high-$t$ Curie-law  limit, with $\bar{\theta}_{\rm 2D}=-1/2$. In 1D, 2D, and 3D, panels (a--c) respectively show that $\chi(t)t$ approaches the asymptotic high-$t$ limits of 2, 1, and 2/3 logarithmically in $t$, linearly in~$t$, and inversely in~$t$, respectively.}
\label{Fig:chi_high_T} 
\end{figure}

It is also of interest to examine the manner in which the 1D, 2D, and 3D susceptibilities asymptote to the Curie-law limit at high temperatures.  In Figs.~\ref{Fig:chi_high_T}(a--c) we plot $\frac{\chi(t)t}{\chi(0)}$ versus $\log_{10}(t)$ for 1D, \mbox{$\chi(t)(t-\bar{\theta}_{\rm 2D})/\chi(0)$} versus $t$ for 2D, and  $\frac{\chi(t)t}{\chi(0)}$ versus $1/t$ for 3D\@.  The choice for the abscissa label for 2D was motivated by the Curie-Weiss behavior in Eq.~(\ref{Eq:chi2DhighT}).  In 1D, the approach of $\chi T$ to its high-$T$ limit was extremely slow, which was found to be logarithmic.  In 3D, the approach of $\chi T$ to its high-$T$ limit was determined by trial and error to vary as $1/T$\@.  The high-$T$ limits are seen from Fig.~\ref{Fig:chi_high_T} to be
\bse
\bea
\lim_{t\to\infty}\frac{\chi(t)t}{\chi(0)} &=& 2 \quad {\rm (1D)}\\
&=& 1  \quad {\rm (2D)}\\
&=& \frac{2}{3}  \quad {\rm (3D)}.
\eea
\ese
Using Eqs.~(\ref{Eqs:chi01D2D3D}) and~(\ref{Eq:chiNC1T}), these results confirm that at high~$T$, the susceptibility in each dimension is just that of $N$ isolated spins.

\section{\label{Sec:Cp} Internal Energy and Heat Capacity at Constant Volume}

The heat capacity $C_{\rm V}$ at constant volume $V$ is defined as
\begin{equation}
C_{\rm V} = \left(\frac{dE_{\rm ave}}{dT}\right)_V,
\end{equation}
where the thermal average (internal) energy per electron is
\begin{equation}
E_{\rm ave}(T) =  \int_0^\infty E\,f(E,T)\,D(E)\,dE.
\end{equation}
At $T = 0$, the Fermi function $f(E,T=0)$ is a step function that is equal to 1 below $E_{\rm F}$ and is equal to 0 above $E_{\rm F}$.  Therefore 
\begin{equation}
E_{\rm ave}(T= 0) = \int_0^{E_{\rm F}} E\,D(E)\,dE.
\label{Eq:Eave(T)}
\end{equation}
Using the density of states $D(E)$ functions in Eqs.~(\ref{Eq:D(EF)2}), one obtains the ground-state energy as
\begin{eqnarray}
E(T= 0) &=& \frac{1}{3} NE_{\rm F} \hspace{0.2in} ({\rm 1D})\nonumber\\*
E(T= 0) &=& \frac{1}{2} NE_{\rm F} \hspace{0.2in} ({\rm 2D})\label{Eq:EaveT=0}\\*
E(T= 0) &=& \frac{3}{5} NE_{\rm F} \hspace{0.2 in} ({\rm 3D}).\nonumber
\end{eqnarray}
The increase in $E_{\rm ave}(T= 0)$ with increasing dimensionality follows because the density of states decreases with energy $E$ in 1D, is constant in 2D and increases with $E$ in 3D\@.

\begin{figure}
\includegraphics[width=3.in]{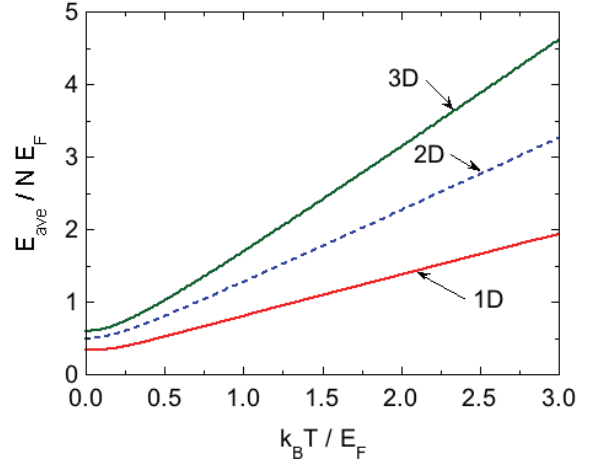}
\caption{Normalized average reduced energy per electron $\epsilon_{\rm ave} = E_{\rm ave}/NE_{\rm F}$ versus reduced temperature $t = k_{\rm B}T/E_{\rm F}$ for free-electron Fermi gases in 1D, 2D, and 3D calculated using Eqs.~(\ref{Eq:e12averages}).}
\label{Fig:Fermi_Gas_1D2D3D_Eave} 
\end{figure}

By rewriting Eq.~(\ref{Eq:Eave(T)}) in terms of the dimensionless variables in Sec.~\ref{Sec:Fund}, one obtains the following expression for the reduced average energy per electron versus reduced temperature
\begin{equation}
\epsilon_{\rm ave}(t) = \int_0^\infty \epsilon\,f(\epsilon,t)\,{\tt D}(\epsilon)\,d\epsilon.
\label{Eq:eave(t)}
\end{equation}
Then using the ${\tt D}(\epsilon)$ functions in Eqs.~(\ref{Eq:D(EF)4}), we obtain 
\begin{eqnarray}
\epsilon_{\rm ave}(t) &=& -\frac{\sqrt{\pi}}{2}\,t^{3/2}\,{\rm Li}_{3/2}\left(-e^{\bar{\mu}/t}\right) \hspace{0.15in}({\rm 1D})\nonumber\\*
\epsilon_{\rm ave}(t) &=& -t^{2}\,{\rm Li}_{2}\left(-e^{\bar{\mu}/t}\right) \hspace{0.65in}({\rm 2D})\label{Eq:e12averages}\\*
\epsilon_{\rm ave}(t) &=& -\frac{9\sqrt{\pi}}{8}\,t^{5/2}\,{\rm Li}_{5/2}\left(-e^{\bar{\mu}/t}\right) \hspace{0.1in}({\rm 3D}),\nonumber
\end{eqnarray}
where the polylogarithm function Li$_n(x)$ was introduced in Eq.~(\ref{Eq:PolyLogFcn}).  In order to plot and/or otherwise utilize these functions, one must first insert the reduced chemical potential $\bar{\mu}(t)$ determined from Eqs.~(\ref{Eq:mu(t)}).  The $\epsilon_{\rm ave}(t)$ functions are plotted in Fig.~\ref{Fig:Fermi_Gas_1D2D3D_Eave} in 1D, 2D, and 3D\@.  The data in Fig.~\ref{Fig:Fermi_Gas_1D2D3D_Eave} for all three dimensions are approximately linear in $T$ for $t\gtrsim 1$, which means that the constant-volume heat capacity $C_{\rm V}(T)=dE_{\rm ave}/dT$ is approximately constant above this temperature.  This is confirmed below.

The $C_{\rm V}(t)$ can be determined by differentiating Eq.~(\ref{Eq:eave(t)}) with respect to $t$, which gives $C_{\rm V}(t)$ per mole of electrons as
\begin{equation}
C_{\rm V}(t) = R\frac{d\epsilon_{\rm ave}(t)}{dt} = R\int_0^\infty \epsilon\,\frac{d f(\epsilon,t)}{d t}\,{\tt D}(\epsilon)\,d\epsilon,
\label{Eq:CV(t)}
\end{equation}
where $R = N_{\rm A}k_{\rm B}$ is the molar gas constant and $N_{\rm A}$ is Avogadro's number.  Keeping in mind that $\bar{\mu}$ is a function of $t$, the total derivative $df/dt$ is obtained as
\begin{eqnarray}
\frac{df(\epsilon,t)}{dt} &=& \frac{1}{t^2}\frac{e^{(\epsilon-\bar{\mu})/t}}{\left[e^{(\epsilon-\bar{\mu})/t} + 1\right]^2}\left[\epsilon - \bar{\mu} + t\frac{d\bar{\mu}(t)}{dt}\right].
\label{Eq:dfdt}
\end{eqnarray}
The temperature derivative $d\bar{\mu}(t)/dt$ is found by taking the total $t$ derivative of Eq.~(\ref{Eq:Findmu}) and then solving for $d\bar{\mu}/dt$, which gives
\begin{equation}
\frac{d\bar{\mu}}{dt}(t) = -\frac{1}{t}\frac{\int_0^\infty \frac{{\tt D}(\epsilon)(\epsilon -\bar{\mu})e^{(\epsilon-\bar{\mu})/t} }{\left[e^{(\epsilon-\bar{\mu})/t}+1\right]^2}d\epsilon}
{\int_0^\infty \frac{{\tt D}(\epsilon)e^{(\epsilon-\bar{\mu})/t} }{\left[e^{(\epsilon-\bar{\mu})/t}+1\right]^2}d\epsilon}.
\label{Eq:dmudt}
\end{equation}
Inserting Eq.~(\ref{Eq:dmudt}) into (\ref{Eq:dfdt}) and the result into~(\ref{Eq:CV(t)}) gives
\bse
\label{Eqs:CV}
\begin{equation}
\frac{C_{\rm V\,1D}}{R} = \frac{\sqrt{\pi t}}{8} \bigg\{\frac{[{\rm Li}_{1/2}(-e^{\bar{\mu}(t)/t})]^2}{{\rm Li}_{-1/2}(-e^{\bar{\mu}(t)/t})}  - 3{\rm Li}_{3/2}\left(-e^{\bar{\mu}(t)/t}\right)  \bigg\},
\end{equation}
\bea
\frac{C_{\rm V\,2D}}{R} &=& -t \left(e^{-\mu(t)/t}+1\right)\left[\ln\left(1+e^{\mu(t)/t}\right)\right]^2\nonumber\\
&& -2t {\rm Li}_2\left(-e^{\mu(t)/t}\right),\\
\frac{C_{\rm V\,3D}}{R} &=& \frac{9\sqrt{\pi}t^{3/2}}{16}\bigg\{ \frac{3\left[{\rm Li}_{3/2}\left(-e^{\mu(t)/t}\right)\right]^2}{{\rm Li}_{1/2}\left(-e^{\mu(t)/t}\right)}\label{Eq:CV3D}\\
&& -5 {\rm Li}_{5/2}\left(-e^{\bar{\mu}(t)/t}\right)\bigg\}.\nonumber
\eea
\ese

\begin{figure}
\includegraphics[width=3.in]{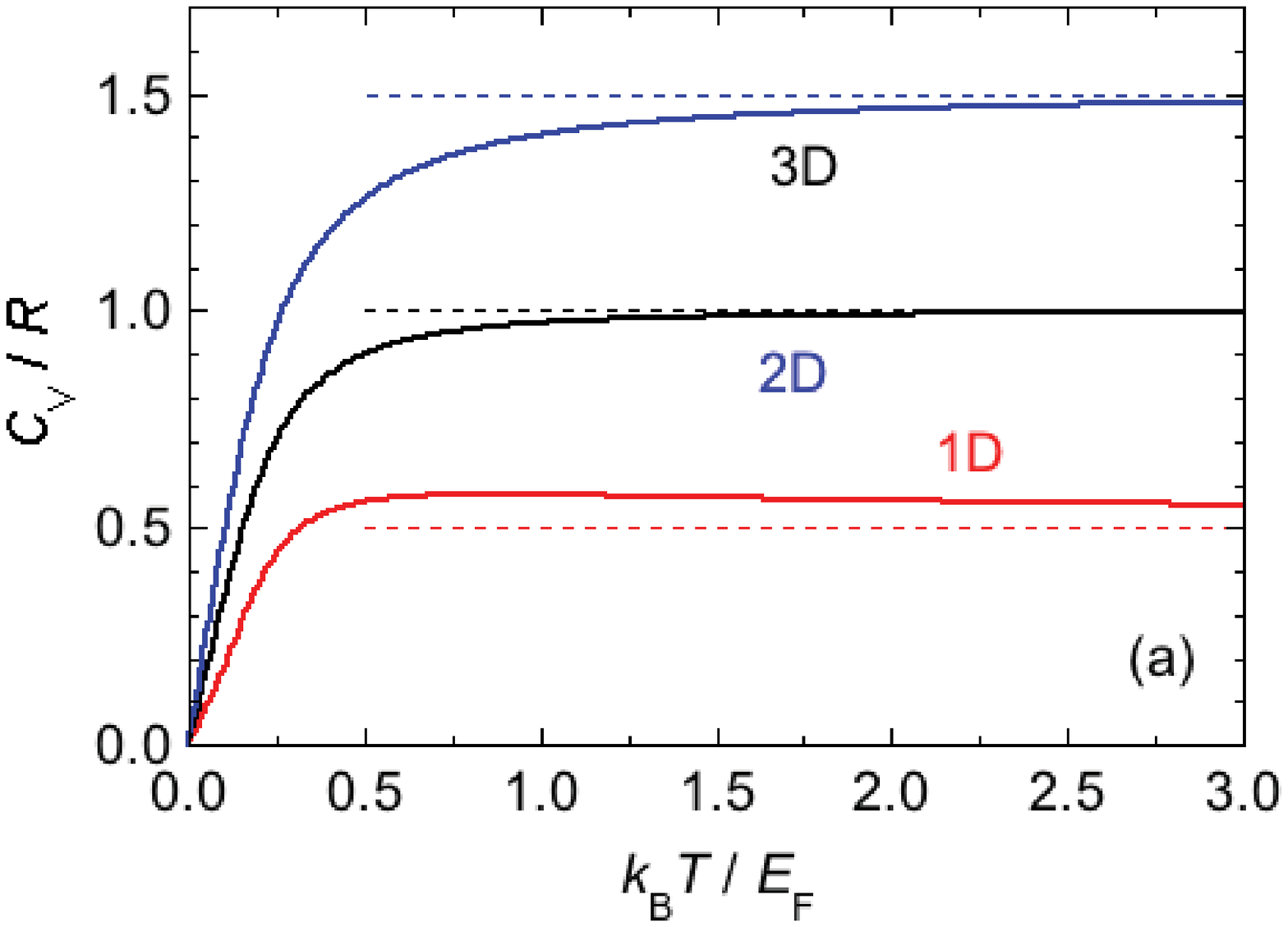}
\includegraphics[width=3.in]{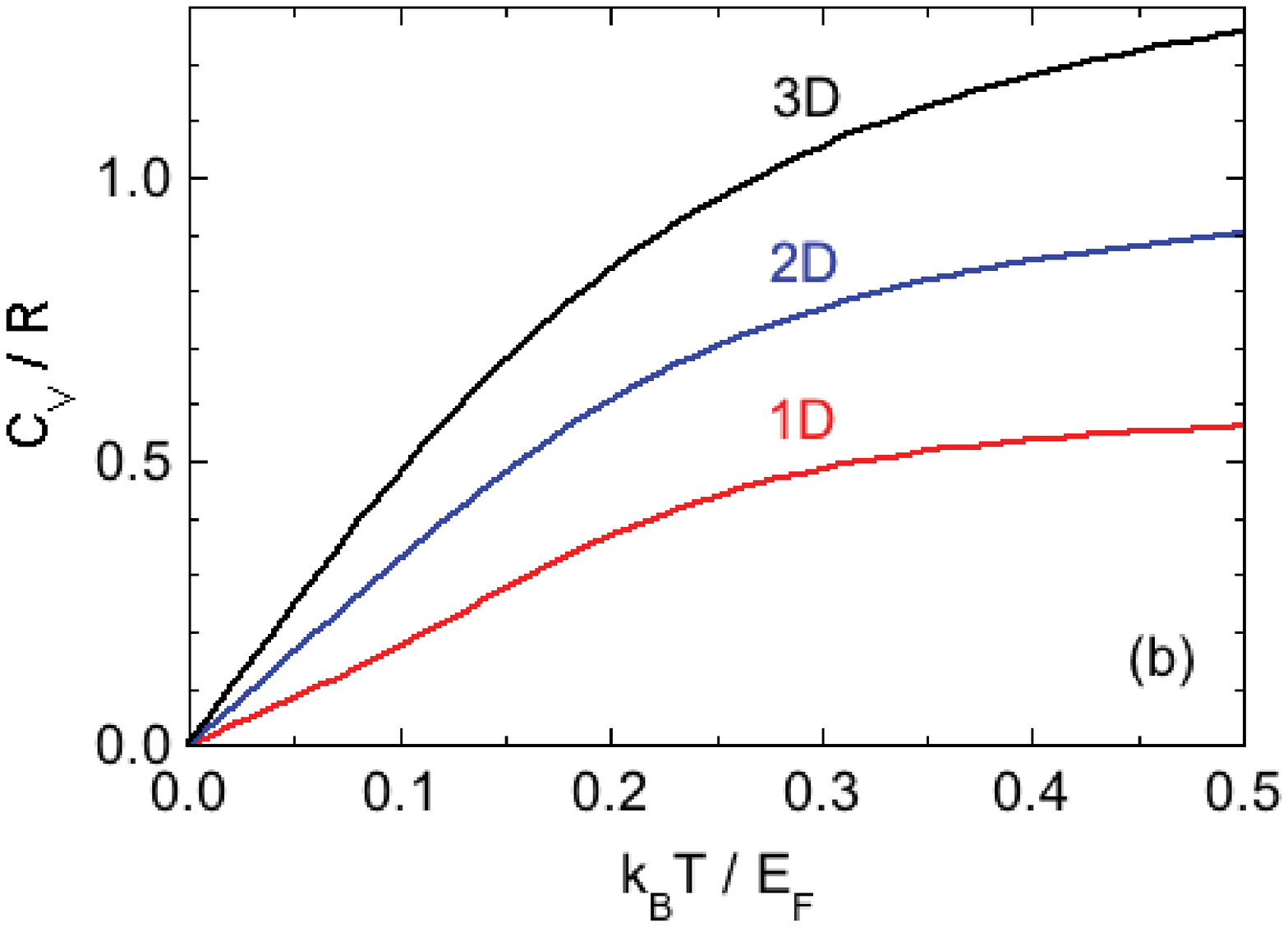}
\caption{(a) Reduced molar heat capacity at constant volume $C_{\rm V}/R$ versus reduced temperature $t = k_{\rm B}T/E_{\rm F}$ for free-electron gases in 1D, 2D, and 3D\@.  The horizontal dashed lines are the predictions $C_{\rm V}/R =  d/2$ for the ideal monatomic gas in 1D, 2D, and 3D, respectively, where $d$ is the dimensionality.  $C_{\rm V}(T)/R$ for the 1D case asymptotically decreases with increasing temperature towards the value $C_{\rm V}/R = 1/2$, in contrast to the monotonic asymptotic increases in the 2D and 3D cases.  (b) Expanded plots of the low-$t$ regions of (a), where $C_{\rm V}\propto T$ at the lowest $T$ for all three dimensions as described by Eq.~(\ref{Eq:CVDE}).}
\label{Fig:Fermi_Gas_1D2D3D_Cv} 
\end{figure}

Using Eqs.~(\ref{Eqs:CV}), $C_{\rm V}(t)$ was calculated for 1D, 2D and 3D $S=1/2$ Fermi gases which are plotted in Fig.~\ref{Fig:Fermi_Gas_1D2D3D_Cv}(a).  At high temperatures, $C_{\rm V}(t)$ approaches the temperature-independent value $C_{\rm V} = Rd/2$ obtained from the equipartition theorem for a classical monatomic ideal gas in the respective dimension $d$ where $R$ is the molar gas constant.  Expanded plots of the low-temperature region are shown in Fig.~\ref{Fig:Fermi_Gas_1D2D3D_Cv}(b), where $C_{\rm V}\propto T$ at the lowest temperatures $t\lesssim 0.05$ for all three dimensions, as discussed next.

Using the ${\tt D}(\epsilon)$ functions in Eqs.~(\ref{Eq:D(EF)4}), the function $g(\epsilon) = \epsilon{\tt D}(\epsilon)$ in Eq.~(\ref{Eq:Int6}) is given by
\begin{eqnarray}
g(\epsilon) &=& \frac{\epsilon^{1/2}}{2}\hspace{0.5in} ({\rm 1D})\nonumber\\*
g(\epsilon) &=& \epsilon\hspace{0.71in} ({\rm 2D})\label{Eq:g(epsilon)}\\*
g(\epsilon) &=& \frac{3\,\epsilon^{3/2}}{2}\hspace{0.41in} ({\rm 3D}).\nonumber
\end{eqnarray}
Substituting these $g(\epsilon)$ functions into Eq.~(\ref{Eq:Int6}) gives
\begin{eqnarray}
\epsilon_{\rm ave}(t) &=& \frac{\bar{\mu}^{3/2}}{3} + \bar{\mu}^{-1/2}\,\frac{\pi^2 t^2}{24}\hspace{0.25in} ({\rm 1D})\nonumber\\*
\epsilon_{\rm ave}(t) &=& \frac{\bar{\mu}^{2}}{2} + \frac{\pi^2 t^2}{6}\hspace{0.74in} ({\rm 2D})\label{Eq:epsilon(mu)}\\*
\epsilon_{\rm ave}(t) &=& \frac{3\bar{\mu}^{5/2}}{5} + \bar{\mu}^{1/2}\,\frac{9\pi^2 t^2}{24}\hspace{0.21in} ({\rm 3D}).\nonumber
\end{eqnarray} 
Inserting the $\bar{\mu}(t)$ dependences in Eqs.~(\ref{Eq:mu1DLoT}), (\ref{Eq:mubar2DLowT}), and~(\ref{Eq:mu3DLoT}) into these expressions, respectively, and using the Taylor-series expansion $(1+x)^n \approx 1+nx$ gives, to order $t^2$,
\begin{eqnarray}
\epsilon_{\rm ave}(t) &=& \frac{1}{3} + \frac{\pi^2 t^2}{12}\hspace{0.25in} ({\rm 1D})\nonumber\\*
\epsilon_{\rm ave}(t) &=& \frac{1}{2} + \frac{\pi^2 t^2}{6} \hspace{0.25in} ({\rm 2D})\label{Eq:epsilon(t)}\\*
\epsilon_{\rm ave}(t) &=& \frac{3}{5} + \frac{\pi^2 t^2}{4} \hspace{0.25in} ({\rm 3D}).\nonumber
\end{eqnarray} 
The values of zero-temperature energy per particle, $\epsilon_{\rm ave}(t=0) = E_{\rm ave}(T=0)/NE_{\rm F}$, agree with the respective values in Eqs.~(\ref{Eq:EaveT=0}).  Finally, using Eq.~(\ref{Eq:CV(t)}) we obtain the low-temperature molar heat capacity expansions
\begin{eqnarray}
C_{\rm V} &=& \frac{\pi^2 R}{6}t\hspace{0.25in} ({\rm 1D})\nonumber\\*
C_{\rm V} &=& \frac{\pi^2 R}{3}t\hspace{0.25in} ({\rm 2D})\label{Eq:CVLowT(t)}\\*
C_{\rm V} &=& \frac{\pi^2 R}{2}t\hspace{0.25in} ({\rm 3D}).\nonumber
\end{eqnarray} 
In terms of the respective conventional densities of states at the Fermi energy $D(E_{\rm F})$ in Eqs.~(\ref{Eq:D(EF)2}), these equations can be written for all three dimensions as
\bse
\begin{equation}
C_{\rm V} = \frac{\pi^2k_{\rm B}^2}{3}D(E_{\rm F})T \equiv \gamma T,
\label{Eq:CVDE}
\end{equation}
where the Sommerfeld coefficient (electronic heat capacity coefficient $\gamma$) is given by the well-known result
\begin{equation}
\gamma = \frac{\pi^2k_{\rm B}^2}{3}D(E_{\rm F}).
\label{Eq:CVgamma}
\end{equation}
\ese
From Fig.~\ref{Fig:Fermi_Gas_1D2D3D_Cv}(b), the linear temperature dependence of $C_{\rm V}$ is seen to be followed for temperatures  \mbox{$T\lesssim 0.05\, T_{\rm F}$}.

\section{\label{Sec:Pressure} Pressure in Two and Three Dimensions}

For a monatomic ideal gas, the equation of state is $pV=Nk_{\rm B}T$, where $p$ is the pressure and $V$ is the volume in 3D and the area in 2D\@.  From the Equipartition Theorem the internal energy is $U=Nk_{\rm B}T$ in 2D and $U=3Nk_{\rm B}T/2$ in 3D.  Thus one has
\bea
pV(T) = \frac{2}{d}U(T),
\label{Eq:pUVIG}
\eea
where $d$ is the dimensionality of the gas.  The same expression is valid at each temperature for the 2D and 3D quantum $S=1/2$ Fermi gases, respectively~\cite{Mancarella2014}.  In reduced units Eq.~(\ref{Eq:pUVIG}) becomes
\bea
\frac{pV}{NE_{\rm F}}(t) &=& \frac{2}{d}\epsilon_{\rm ave}(t).
\label{Eq:pUVIG2}
\eea

Using the second and third of Eqs.~(\ref{Eq:e12averages}), we obtain
\bse
\label{Eqs:pVNEF}
\bea
\frac{pV}{NE_{\rm F}} &=& -t^2 {\rm Li}_2(-e^{\bar{\mu}(t)/t}) \quad ({\rm 2D}),\\
\frac{pV}{NE_{\rm F}} &=& -\frac{3\sqrt{\pi}}{4}\,t^{5/2}\,{\rm Li}_{5/2}(-e^{\bar{\mu}(t)/t})	\quad ({\rm 3D}),
\eea
\ese
where $\bar{\mu}(t)$ is calculated by solving the second and third of Eqs.~(\ref{Eq:FindMu}), respectively.  The compression factors obtained from Eqs.~(\ref{Eqs:pVNEF}) are
\bse
\label{Eq:CompFactor}
\bea
\frac{pV}{Nk_{\rm B}T} &=& -t {\rm Li}_2(-e^{\bar{\mu}(t)/t}) \ \  ({\rm 2D}),\\
\frac{pV}{Nk_{\rm B}T} &=&  -\frac{3\sqrt{\pi}}{4}\,t^{3/2}\,{\rm Li}_{5/2}(-e^{\bar{\mu}(t)/t})\ {(\rm 3D)}.
\eea
\ese

\begin{figure}[t]
\includegraphics[width=3.in]{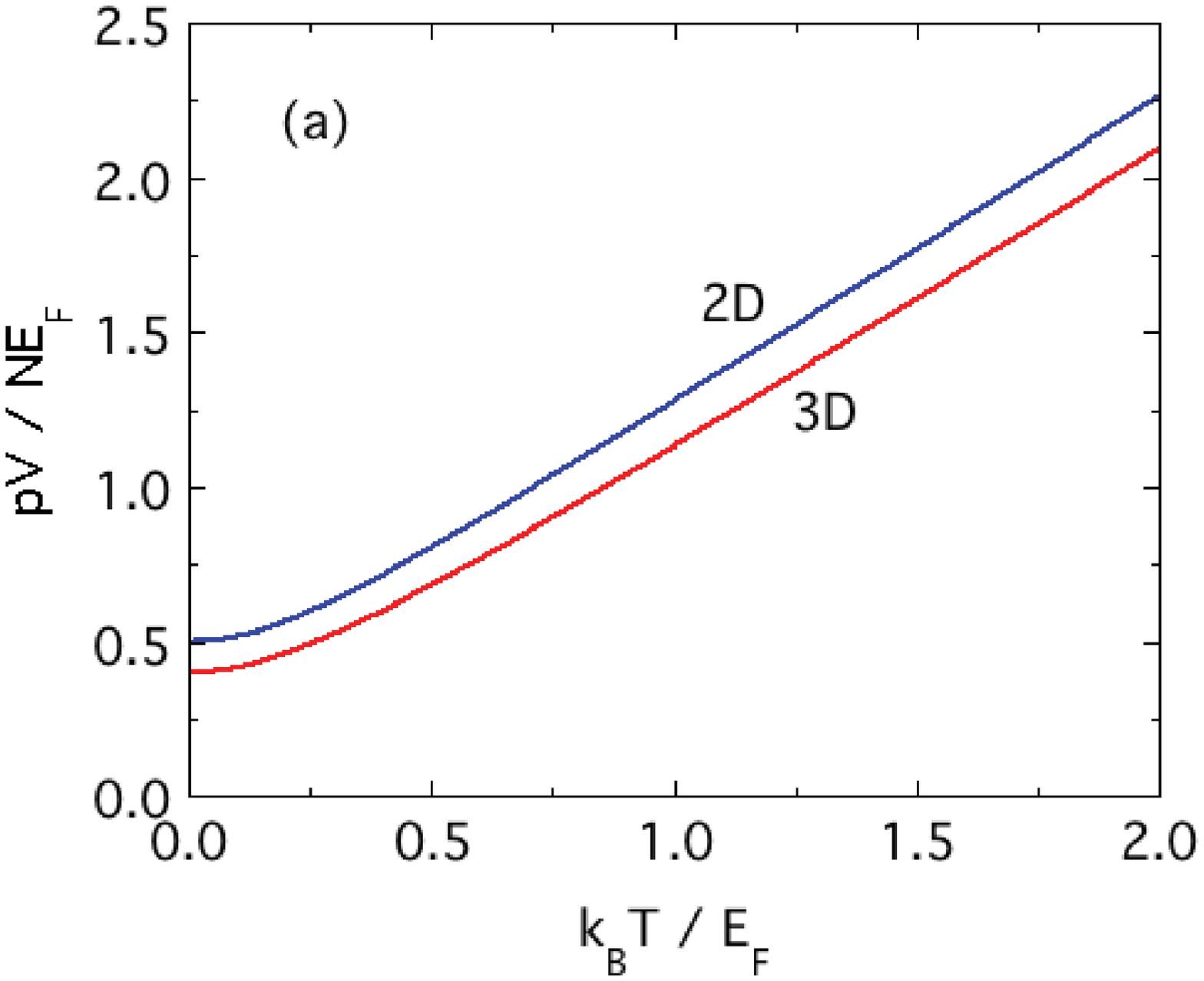}
\includegraphics[width=3.in]{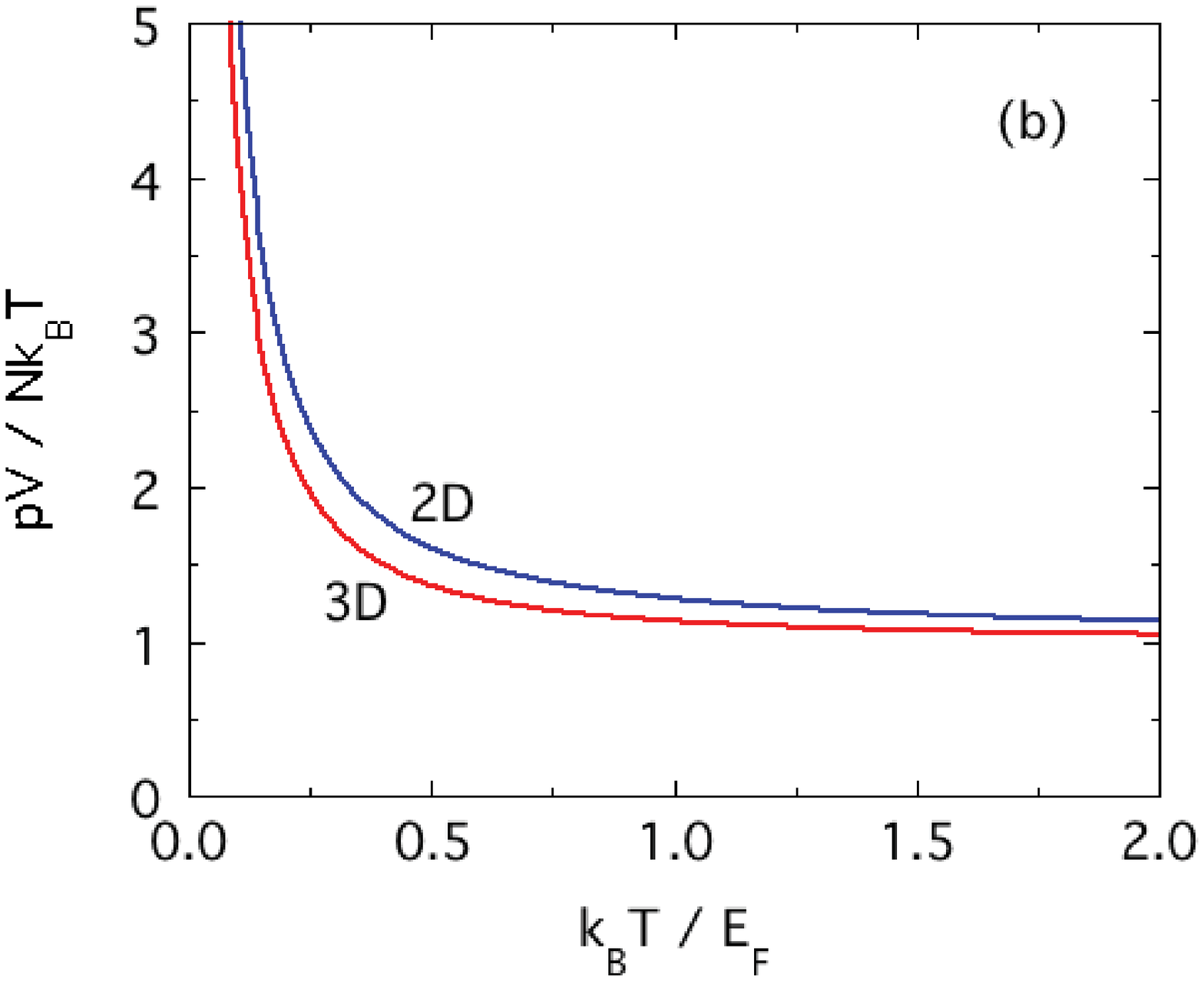}
\caption{(a) Reduced pressure $pV/NE_{\rm F}$ versus reduced temperature $t=k_{\rm B}T/E_{\rm F}$ for a free-electron Fermi gas in three dimensions.  (b)~Compression factor $pV/Nk_{\rm B}T$ versus~$t$ for the same gas.  For 2D, $V$ is the area of the gas.}
\label{Fig:pVonEF_pVonkT} 
\end{figure}

The scaled pressures $pV/NE_{\rm F}$ versus reduced temperature $t=k_{\rm B}T/E_{\rm F}$ computed for 2D and 3D electron Fermi gases using Eqs.~(\ref{Eqs:pVNEF}) are plotted in Fig.~\ref{Fig:pVonEF_pVonkT}(a)~\cite{Wiki}.  As seen in the figure, the reduced degeneracy pressure is $pV/NE_{\rm F}=0.4$ at $t\to0$ in 3D and 0.5 in 2D\@.  The pressure increases with increasing temperature, approaching a classical proportional behavior at high temperatures.  This behavior is examined in more detail in Fig.~\ref{Fig:pVonEF_pVonkT}(b), where the compression factors $pV/Nk_{\rm B}T$ are plotted versus $t$ using Eqs.~(\ref{Eq:CompFactor}).  At low~$t$, this quantity diverges as $1/t$ because $p$ attains a constant value for $t\to0$ as shown in Fig.~\ref{Fig:pVonEF_pVonkT}(a).  At high~$t$, the compression factors approach unity since the high-$t$ limit corresponds to the ideal-gas law.

\section{Enthalpy and Heat Capacity at Constant Pressure}

\begin{figure}
\includegraphics[width=3.in]{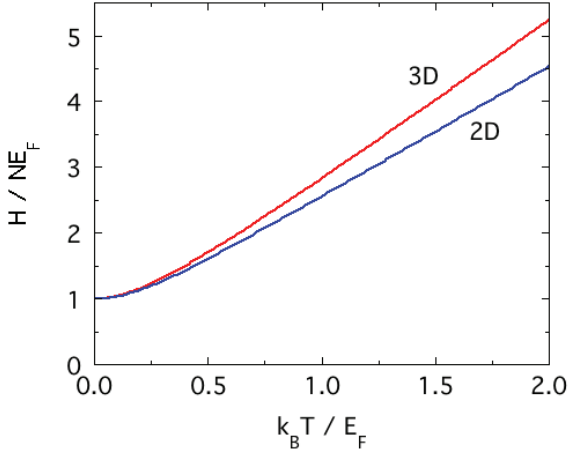}
\caption{Reduced enthalpy $H/NE_{\rm F}$ versus reduced temperature $t=k_{\rm B}T/E_{\rm F}$ for a $S-1/2$ Fermi gas in 2D and 3D.}
\label{Fig:Enthalpy_2D3D} 
\end{figure}

Here only the 2D and 3D cases are relevant.  The enthalpy $H$ is defined as
\bea
H = U + pV.
\eea
In reduced variables, this expression becomes
\bea
\frac{H}{NE_{\rm F}}(t) = \epsilon_{\rm ave}(t)  + \frac{pV}{NE_{\rm F}}(t),
\label{Eqs:H}
\eea
where the temperature dependences of the first and second term are given in Eqs.~(\ref{Eq:e12averages}) and~(\ref{Eqs:pVNEF})  and plotted for 2D and 3D  in Figs.~\ref{Fig:Fermi_Gas_1D2D3D_Eave}  and~\ref{Fig:pVonEF_pVonkT}(a), respectively.   Plots of the reduced enthalpy versus reduced temperature in 2D and 3D are shown in Fig.~\ref{Fig:Enthalpy_2D3D}.

The heat capacity at constant pressure is given by $C_{\rm p}=(\partial H/\partial T)$.  This yields the molar heat capacity as
\bea
\frac{C_{\rm p}}{R} = \frac{\partial [H/(NE_{\rm F})]}{\partial t}.
\eea 
Then using Eq.~(\ref{Eqs:H}) one obtains
\bea
\frac{C_{\rm p}(t)}{R} = \frac{C_{\rm V}(t)}{R} + \frac{\partial}{\partial t}\left(\frac{pV}{NE_{\rm F}}(t)\right),
\eea
where the first and second terms are calculated from Eqs.~(\ref{Eqs:CV}) and Eqs.~(\ref{Eqs:pVNEF}), respectively.  At all temperatures we find the exact results
\bse
\label{Eqs:CpCV}
\bea
 \frac{C_{\rm p}(t)}{C_{\rm V}(t)} &=& 2 \qquad (2{\rm D}),\\
\frac{C_{\rm p}(t)}{C_{\rm V}(t)} &=& \frac{5}{3} \qquad (3{\rm D}),
\eea
\ese
where $C_{\rm V}(t)$ was calculated for both dimensions in Sec.~\ref{Sec:Cp} and plotted in 1D, 2D, and 3D in Fig.~\ref{Fig:Fermi_Gas_1D2D3D_Cv}.  Interestingly, the results in Eqs.~(\ref{Eqs:CpCV}) are the same as for a monatomic ideal gas in 2D and~3D, respectively.

\section{\label{Sec:Compress} Isothermal Bulk Modulus}

The isothermal bulk modulus $K_T$ is defined as 
\bea
K_T = -V\left(\frac{dp}{dV}\right)_T,
\label{Eq:KTDef}
\eea
with dimensions of pressure.  Figure~\ref{Fig:pVonEF_pVonkT}(b) plotted $pV/Nk_{\rm B}T$ versus $t=k_{\rm B}T/E_{\rm F}$.  Thus at fixed $T$ one has 
\bse
\bea
\frac{pV}{Nk_{\rm B}T} &=& f(t)\ ({\rm a\ number}), \\
p &=& \frac{Nk_{\rm B}T}{V}f(t), \\
\left(\frac{\partial p}{\partial V}\right)_T &=& -\frac{p}{V}.
\eea
\ese
Hence using Eq.~(\ref{Eq:KTDef}) one obtains for 2D and 3D free-electron Fermi gases the same expression
\bse
\bea
K_T(t) &=& p(t),\label{Eq:KTcalc}\\
\frac{K_TV}{NE_{\rm F}} &=& \frac{pV}{NE_{\rm F}}.
\eea
\ese
Equation~(\ref{Eq:KTcalc}) is the same as for a monatomic ideal gas in either 2D or 3D\@.  The reduced pressure $pV/NE_{\rm F}$ was plotted versus $t$ for 2D and 3D free-electron Fermi gases in Fig.~\ref{Fig:pVonEF_pVonkT}(a). 

\section{\label{ThermExpand} Thermal Expansion Coefficient}

\begin{figure}
\includegraphics[width=3.in]{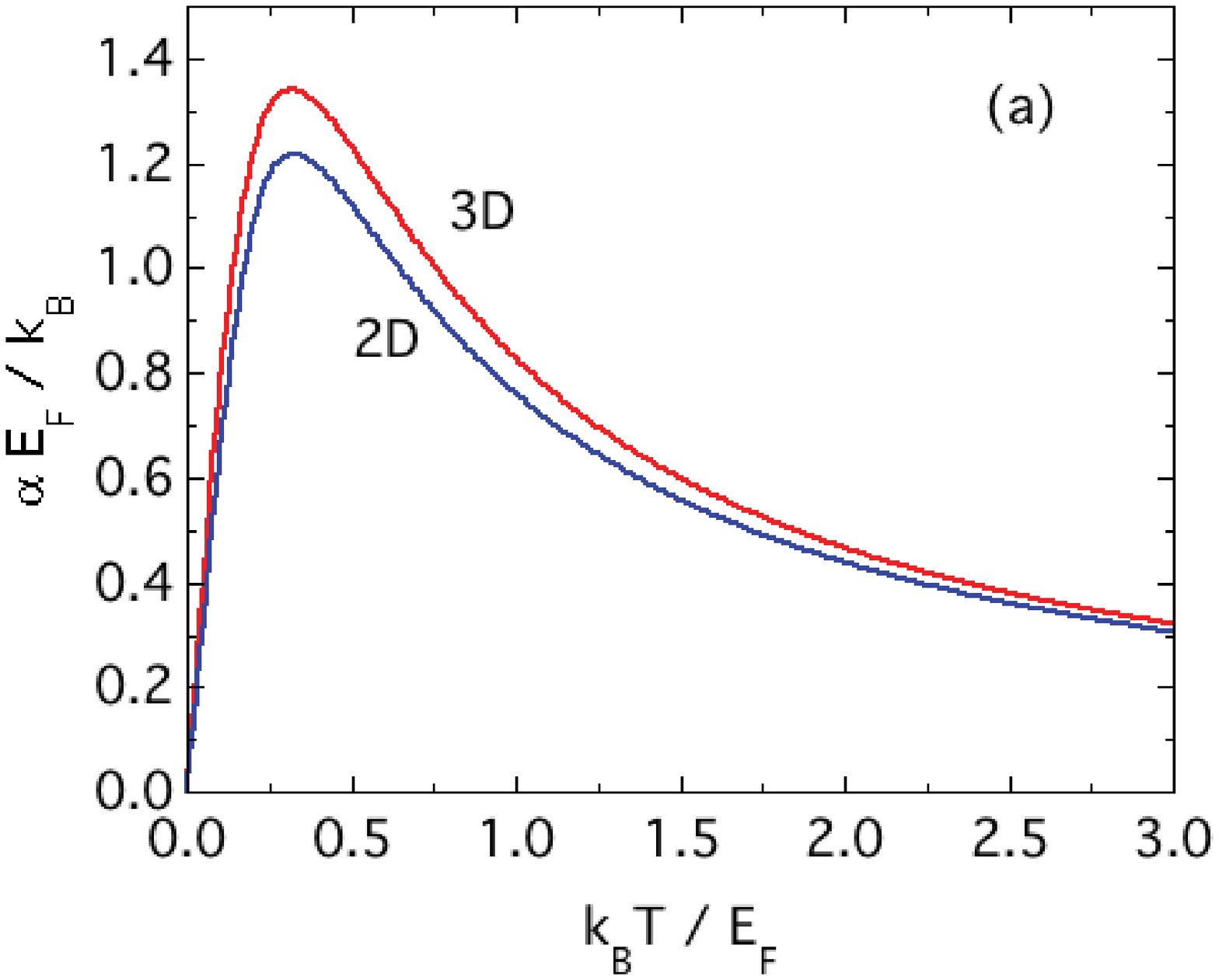}
\includegraphics[width=3.in]{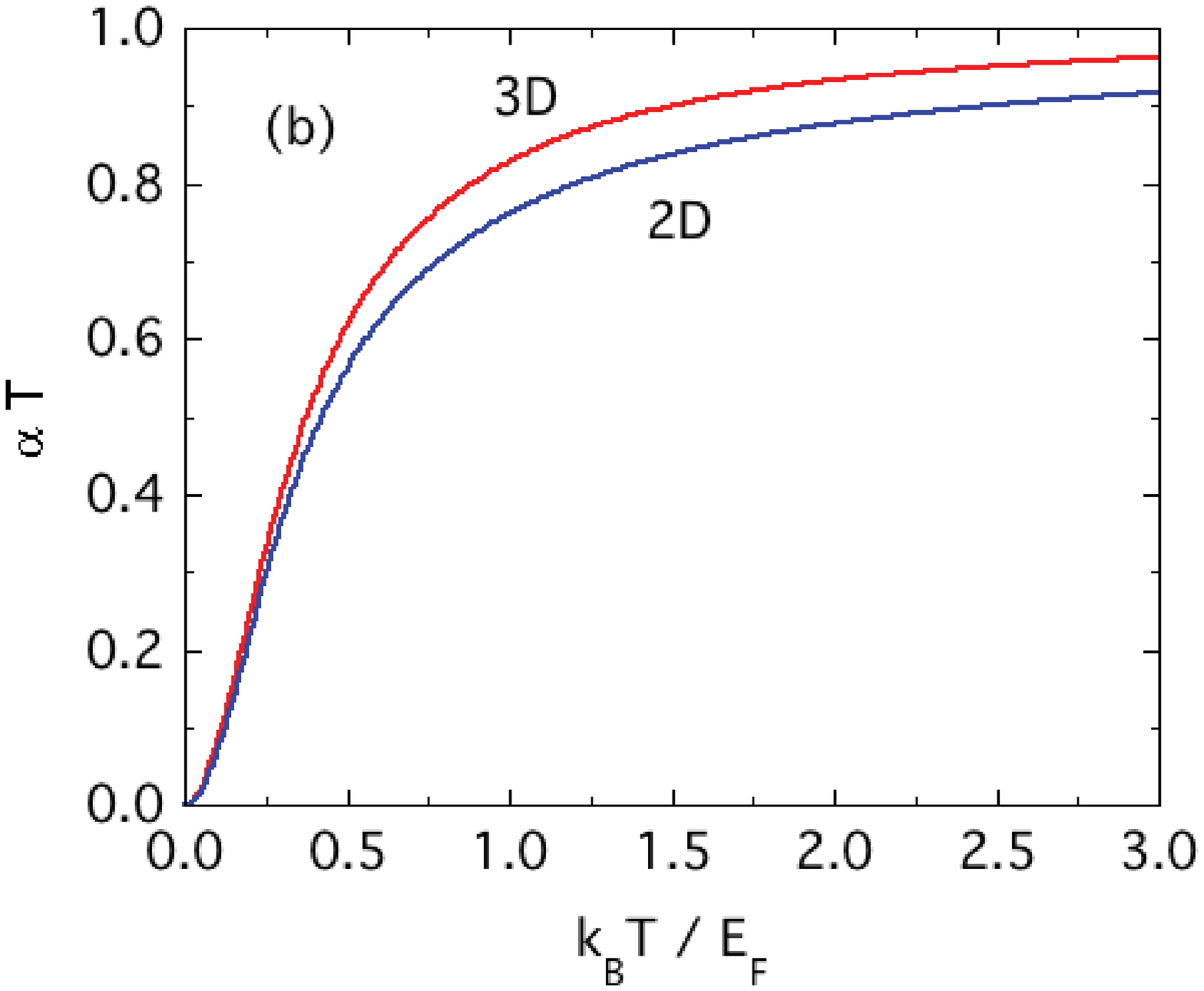}
\caption{(a)~Reduced thermal expansion coefficient~$\alpha E_{\rm F}/k_{\rm B}$ versus  reduced temperature $t = k_{\rm B}T/E_{\rm F}$  for a free-electron Fermi gas in 2D and 3D\@. (b)~$\alpha T$ versus~$t$.  The data approach unity at high temperatures, which is the property of a monatomic ideal gas.}
\label{Fig:alpha_2D3D} 
\end{figure}

The thermal expansion coefficient $\alpha$ is defined as 
\bea
\alpha = \frac{1}{V}\left(\frac{\partial V}{\partial T}\right)_p.
\eea
A useful Maxwell relation is
\bea
\alpha = \frac{1}{K_T}\left(\frac{\partial p}{\partial T}\right)_V,
\eea
where for ideal gases and $S=1/2$ Fermi gases $K_T=p$ from Eq.~(\ref{Eq:KTcalc}).  Therefore
\bea
\alpha = \frac{1}{p} \left(\frac{\partial p}{\partial T}\right)_V.
\eea
The reduced pressure and temperature are
\bea
\bar{p}= \frac{pV}{NE_{\rm F}},\qquad t = \frac{k_{\rm B}T}{NE_{\rm F}},
\eea
so the reduced thermal expansion coefficient is
\bea
\frac{\alpha E_{\rm F}}{k_{\rm B}} = \frac{1}{\bar{p}}\left(\frac{\partial \bar{p}}{\partial t}\right)_V,
\eea
where $\bar{p}(t)$ was calculated in Sec.~\ref{Sec:Pressure}.  In terms of the $T$-dependent chemical potential, we obtain 
\bea
\frac{\alpha E_{\rm F}}{k_{\rm B}} =  \frac{1}{2t}\left\{5 - \frac{3\left[{\rm Li}_{3/2}(-e^{\bar{\mu}(t)/t})\right]^2}{{\rm Li}_{1/2}(-e^{\bar{\mu}(t)/t}){\rm Li}_{5/2}(-e^{\bar{\mu}(t)/t})} \right\}.
\eea

Figure~\ref{Fig:alpha_2D3D}(a) shows plots of $\alpha E_{\rm F}/k_{\rm B}$ versus~$t$ for 2D and 3D $S=1/2$ Fermi gases, which show that {\mbox{$\alpha(t=0)=0$}.  The curves show peaks at $t \approx 0.3$, followed by a smooth decrease at higher temperatures.  To examine this decrease quantitatively, $\alpha T$ versus~$t$ is plotted for the two dimensions  in Fig.~\ref{Fig:alpha_2D3D}(b).  The data for both dimensions asymptotically approach unity, which is the value for the ideal gas since $\alpha = 1/T$ for that gas.

\section{\label{ConclRem} Concluding Remarks}

The properties calculated versus dimension and temperature in this paper include the magnetic spin susceptibility, internal energy, heat capacity at constant volume, pressure, enthalpy, heat capacity at constant pressure, isothermal bulk modulus, and thermal-expansion coefficient.  The crossovers of these properties  between the low-temperature degenerate regime and the high-temperature nondegenerate regime were elucidated and analyzed.  The high-temperature limit of the magnetic susceptibility is that of isolated spins-1/2, and those of the other thermodynamic quantities correspond to the respective properties of the monatomic ideal gas.

The expressions for three of the thermodynamic properties have forms analogous to those for the monatomic ideal gas.  The temperature-independent ratio of heat capacities $C_{\rm p}/C_{\rm V}$ for the free-electron Fermi gas in 2D and 3D in Eqs.~(\ref{Eqs:CpCV}), the temperature-dependent expressions for the pressure times the volume in terms of the internal energy for the respective dimension, and in each of 2 and 3 dimensions the expression~(\ref{Eq:KTcalc}) for the isothermal bulk modulus for both 2D and 3D electron Fermi gases are the same as for the monatomic ideal gas.

Usually the theory for the nonrelativistic free-electron Fermi gas treated here is only of practical interest at temperatures low compared to the Fermi temperature, because the latter temperature for most metals is very high compared to the maximum accessible measurement temperature.  However one can envision metals with low Fermi temperatures for which the present results might be useful.  Of particular interest might be semimetals with low carrier concentrations.  The challenge would be to separate the properties of the Fermi gas from those of the lattice.  However, the magnetic properties of a Fermi gas might be measurable from the degenerate to the nondegenerate regime if the purity of the nonmagnetic host was sufficient.  A special case is the $S=1/2$ nuclear magnetic susceptibility of liquid $^3$He, for which the experimental data from the literature for the magnetic susceptibility versus temperature at a pressure of 20.5~atm  in Fig.~\ref{3He_chiFit} were fitted well by the theory from a temperature of 0.045~K in the degenerate temperature regime to 1~K in the nondegenerate temperature regime, yielding a Fermi temperature of 0.30~K\@.

\acknowledgments

I am grateful to H. Godfrin, R. J. McQueeney, and {\mbox{D. Vaknin} for helpful comments.  This research was partially supported by the U.S. Department of Energy, Office of Basic Energy Sciences, Division of Materials Sciences and Engineering.  Ames Laboratory is operated for the U.S. Department of Energy by Iowa State University under Contract No.~DE-AC02-07CH11358.


\begin{thebibliography}{20}
\bibitem{Mathematica} All calculations in this paper were carried out using the program {\tt Mathematica}, in which the polylogarithm functions are built in.  Polylogarithm functions are solutions of Complete Fermi-Dirac integrals.
\bibitem{Wiki} Qualitative plots similar to Fig.~\ref{Fig:pVonEF_pVonkT}(a) but with no precise axis labels and no numerical axis scales have appeared in several Wikipedia entries.
\end{thebibliography}
\end{document}